\documentclass[lettersize,journal]{IEEEtran}
\usepackage{amsmath,amsfonts}
\usepackage[utf8]{inputenc}
\usepackage{algorithmic}
\usepackage{algorithm}
\usepackage{array}
\usepackage[caption=false,font=footnotesize,labelfont=rm,textfont=rm]{subfig}
\usepackage{textcomp}
\usepackage{stfloats}
\usepackage{url}
\usepackage{multirow}
\usepackage{verbatim}
\usepackage{graphicx}
\usepackage{cite}
\usepackage{soul}
\usepackage{color}
\usepackage{booktabs}

\usepackage{xcolor}
\def\BibTeX{{\rm B\kern-.05em{\sc i\kern-.025em b}\kern-.08em
		T\kern-.1667em\lower.7ex\hbox{E}\kern-.125emX}}
\usepackage{balance}
\begin{document}
	
	\title{Few-Shot Specific Emitter Identification via Integrated Complex Variational Mode Decomposition and Spatial Attention Transfer}
	
	\author{\IEEEauthorblockN{Chenyu Zhu, Zeyang Li,
			Ziyi Xie, Jie Zhang}
		\thanks{
			The e-mail: 24B952023@stu.hit.edu.cn.
			
			The related codes will be updated on GitHub gradually: https://github.com/ZhuCY-loong00/ICVMD-SAT
	}}

	\markboth{}%
	{Shell \MakeLowercase{\textit{et al.}}: A Sample Article Using IEEEtran.cls for IEEE Journals}

	\maketitle
	
	\begin{abstract}
		Specific emitter identification (SEI) utilizes passive hardware characteristics to authenticate transmitters, providing an effective physical-layer security solution. However, most deep-learning-based methods rely on extensive data or require prior information, which poses challenges in real-world scenarios with limited labeled data. We propose an integrated complex variational mode decomposition algorithm that decomposes and reconstructs complex-valued signals to approximate the original transmitted signals, thereby enabling more accurate feature extraction. A multi-scale temporal convolutional network is utilized to effectively model the sequential signal characteristics. We introduce a signal-guided spatial attention mechanism to adaptively weight informative signal segments, which enhances identification performance. The branch attention network allows leveraging pre-trained weights from other data, and it reduces the requirements for auxiliary datasets. Ablation experiments on simulated data demonstrate the effectiveness of each component in the model. On a public dataset, our method achieves at least 96\% accuracy for 30 emitters using only 10 symbols without requiring task-specific signal information or emitters' prior knowledge.
	\end{abstract}
	
	\section{Introduction}
	\IEEEPARstart{W}{ith} rapid advances in wireless communication technologies, numerous devices have been deployed and interconnected in large-scale Internet of Things (IoT) and cognitive radio networks \cite{iot}. As vast amounts of sensitive data and vital information are transmitted wirelessly, secure and reliable wireless equipment authentication has become increasingly critical. Conventional security methods, such as address-based access control and cryptographic algorithms, face severe vulnerabilities due to potential key leakage and hardware spoofing \cite{tit}. Various resource-constrained devices in the network further complicate security challenges, while distributed key management incurs additional operational costs. These challenges drive the need for more reliable and lightweight security mechanisms, leading to the introduction of radio frequency fingerprint (RFF) within the specific emitter identification (SEI) field. As a passive technique, RFF identification (RFFI) relies entirely on received signals, making it suitable for low-cost device management \cite{jstsp}.
	
	SEI identifies transmitters by extracting distinctive features, making feature selection and extraction its core. Early approaches relied on intentional modulation (IM) parameters, which limits their ability to distinguish emitters with multifunction capabilities or similar characteristics \cite{proieee}. Subsequent research revealed that hardware variations arising from manufacturing processes lead to unavoidable device differences, even among those from the same batch \cite{sensorj}. These unique hardware-induced unintentional modulation (UIM) characteristics are difficult to forge, providing robust passive physical-layer authentication \cite{pro}. Thus, RFFI is in practice treated as synonymous with SEI due to its prevalent application \cite{rff976}.
	
	Appropriate signal processing generally improves RFF extraction. Early studies employed various time-frequency transformation techniques, such as the short-time Fourier transform, Hilbert-Huang transform (HHT), and wavelet transform, to convert signals into images and utilize them for subsequent classification \cite{STFT,HHT,wavelet}. However, these transformations may discard critical signal information that compromises the ability to differentiate between similar transmitters, and existing studies lack consensus on the optimal transformation for each IM type \cite{tsp}. 
	
	Sequence-based methods have been increasingly adopted because they preserve complete temporal dependencies and provide comprehensive information \cite{wang}. Many studies have adopted raw in-phase/quadrature (I/Q) samples as inputs. Recently, data-driven signal decomposition methods, such as variational mode decomposition (VMD), have also been increasingly employed to enhance RFFI performance. VMD can decompose modes non-recursively and simultaneously in the frequency domain, simplifying subsequent signal analysis or feature extraction for RFFI applications\cite{vmd3}. For instance, \cite{vmd1} introduced a SEI scheme based on VMD and spectral features. In \cite{vmd2}, adaptive signal decomposition was performed via bivariate VMD for denoising, proving suitable for low signal-to-noise ratio (SNR) scenarios.
	
	However, conventional RFF extraction approaches exhibit significant drawbacks. Extracting handcrafted features generally requires experience and domain knowledge, which limit these methods' applicability in diverse application scenarios. These inherent flaws have spurred interest in data-driven deep learning approaches, as they avoid complex manual feature engineering. Neural networks (NN), initially employed solely as classifiers, can autonomously extract potential non-linear high-dimensional features. Research indicates that end-to-end methods with large-scale labeled data outperform those entirely based on features and classifiers \cite{end2end}.
	
	Although recent research has achieved breakthroughs in various controlled settings, most approaches treat SEI as a supervised classification task under the assumption of abundant labeled samples. Real-world scenarios are more complicated, and captured samples are often limited in both quantity and quality, leading to open-set, few-shot (FS), dynamic noise ranges, and variable-modulation identification issues \cite{open,dagu2, dtSNR,biantiaozhi}. Among these, the FS-SEI problem is the most common and fundamental, as insufficient training samples reduce feature diversity and further exacerbate other issues. Addressing FS-SEI thus provides a foundation for solving broader challenges.
	\subsection{Related Works}
	While NNs typically require extensive labeled data for training, insufficient samples can degrade model efficacy and even prevent the construction of effective models. Research on FS-SEI typically focuses on mitigating sample deficiency impacts through four principal approaches:
	
	\subsubsection{Feature Extraction} 
	Given sufficient prior knowledge, such as signal parameters and detailed channel information, it becomes possible to process signals in a targeted manner and acquire distinctive features. Although these methods may not be designed explicitly for FS settings, they surpass traditional ones and maintain performance under limited data conditions.
	
	The SRP-CBL model in \cite{SRP} converts signals into 2D recurrence plots (SRPs) and employs a convolutional broad learning (CBL) framework to extract intrinsic patterns. It reduces computational complexity and yields superior accuracy in limited-data scenarios. By mining embedded temporal information, Hu \textit{et al.} introduced a representation approach to visualize RFFs and enhance data separability \cite{buwanmei}. They also proposed a technique to quickly estimate and compensate the carrier frequency offset for cyclic shift signals \cite{cycle}. This technique improves the noise resistance of RFFs, and a simple classifier increases testing speed. An accuracy of 96.66\% is achieved in line-of-sight (LOS) scenarios with only 10 symbols. Unfortunately, these approaches are specific and require much difficult-to-obtain information, posing challenges for practical model construction.
	\subsubsection{Transfer Learning}
	When target samples are limited, we can retrieve similar ones from databases as supplements. Transfer learning aims to build a generalizable model in the source domain and then apply this knowledge to a different yet interrelated target domain. It improves outcomes while reducing reliance on extensive target data.
	
	The zero-shot method in \cite{acm} identifies robust features effective for small-sample scenarios by excluding various data in the auxiliary dataset. \cite{probe} transforms signals in various ways, then extracts probe-features from the obtained time-frequency representations, which improves recognition accuracy under scarce data conditions. \cite{AMAE} uses unlabeled training samples to train an asymmetric mask self-encoder, obtaining a good RFF extractor. The classifier is then fine-tuned using a small number of labeled samples. \cite{ADSB} employs a zero-padding preprocessing algorithm to mitigate input ambient noise and interference, and the RFF extractor's likelihood of overfitting is reduced through a Siamese network.
	
	However, most transfer methods are feature-specific and need suitable auxiliary data. The effectiveness depends on the relevance of domains, making it difficult to establish connections between weakly related data and rendering the knowledge ineffective. When the domain gap is large (e.g., cross-modulation types), negative transfer may occur.
	\subsubsection{Data Augmentation}
	We can choose to imitate and generate synthetic samples based on existing data to alleviate data scarcity. Some techniques like slicing, flipping, and adding noise are used in \cite{conf}, but they fail to produce meaningful new data, resulting in limited performance. \cite{wcl} extracts key data from existing samples and generates signals by masked autoencoding techniques. NNs are pre-trained by predicting masked signal segments, requiring no additional labeled data. Alternatively, \cite{IntML} introduces InterML, which expands the sample scale via interpolation in the sample space and designs a specialized metric to increase feature discrimination in the feature space.
	
	Current methods are often based on generative adversarial networks (GANs), which operate by training a pair of competing networks, one generating synthetic signals, and the other differentiating them from real signals \cite{gan2}. The discriminator provides feedback to help the generator. When the model approaches convergence, the generated signals closely resemble actual signals. \cite{dagu2} utilizes the contour stella image for feature visualization, and the signals by GAN are filtered by structural similarity thereby improving their quality. The semi-supervised method in \cite{semi} employs a confidence threshold to pseudo-label some weakly augmented unlabeled samples, and uses them as annotations for other samples. This operation enhances feature consistency among samples of the same category, and increases diversity by using deep metric learning.
	
	However, these methods rely entirely on adversarial processes, and the generator does not have direct access to instances. As subtle distortions, RFFs cannot be easily visually identified like images, causing it challenging to discern whether generated signals carry the appropriate features. No methods can deduce a complete distribution from limited samples, resulting in signals lacking completeness and diversity, and various implementations under the same principle yield entirely different outcomes. Adapting these models to new datasets frequently fails to achieve the expected results, along with issues like training instability and model collapse, making them sensitive to training skills \cite{gan3}.
	\subsubsection{Meta-Learning}
	Meta-learning focuses on enhancing learning abilities by training across various scenarios, enabling automatic and rapid parameter tuning for new tasks. Meta-learners integrate experiences from various base learning tasks to capture commonalities, thereby reducing iterations in practical tasks. Some studies have introduced it to address model instability due to insufficient data.
	
	The model-agnostic meta-learning (MAML) approach proposed in \cite{MAML} achieves high accuracy and broad applicability across various emitter types using only a few labeled samples. MAML is compatible with any model using gradient descent and is particularly suited for fine-tuning. The method presented in \cite{sensorj} processes signals with HHT before encoding, then the hidden features extraction and classification network is trained with meta-learning in FS settings. \cite{access} preprocesses signals with bispectrum and Radon transform, extracting features and calculating the distances and divergences. All distances are normalized to facilitate meta-learning effectively.
	
	Although both meta-learning and transfer learning use extra data, their focus differ. Transfer learning emphasizes obtaining distinguishable features, while meta-learning focuses on NNs, aiming to train networks with generalization capabilities using sufficient samples. Transfer learning often applies pre-trained models directly to new tasks, whereas meta-learning enables quick adaptation to new settings with minimal data because it has constructed numerous tasks requiring high training costs \cite{wcl}.
	\subsection{Motivations and Contributions}
	To mitigate reliance on prior knowledge and high-quality samples in the aforementioned methods, we construct a novel framework for FS-SEI. RFFs, as UIM, can be viewed as parasitic distortions carried within signals. Previous studies generally consider only extracting features from the received signal while ignoring their correspondence with the original signals. However, relying solely on these features is often insufficient for robust classification in FS scenarios. If the original signal could be obtained simultaneously to establish the connection, it would enable accurate models even with minimal data, which is high-cost and practically impossible in non-cooperative scenarios. Although \cite{blind} attempts to decompose UIM and IM components adaptively, its data-driven framework requires substantial training data. While traditional VMD is limited to real-valued signals, the existing complex extension proposed in \cite{CVMD}, is designed as a generic tool that is not adapted for specific signal characteristics. Therefore, we propose an integrated complex variational mode decomposition (ICVMD) algorithm, aiming to reconstruct an approximation or distinctive representation of the original signal by leveraging frequency properties.
	
	Fully connected (FC) classifiers in traditional CNNs are prone to overfitting with high parameter count but few training samples, causing discrepancies when the signal segment shifts in position. Consequently, our proposed model utilizes a fully convolutional network (FCN) architecture and incorporates a spatial attention mechanism (SAM) to adaptively adjust each signal segment's feature weight. The contributions are summarized as follows:
	\begin{itemize}
		\item We propose an SEI-adapted extension of traditional VMD to the complex domain, enabling meaningful and consistent representation of original signals to extract emitter-specific features. The feature transfer capability is also improved by constructing the association of the decomposed signals. 
		\item To better exploit temporal correlations within signal sequences, we employ a temporal convolutional network (TCN) to analyze the extracted feature sequences. The TCN provides improved sequential analysis and robustness in feature extraction over conventional methods. By utilizing various dilation factors as adaptive band-pass filters, the network captures multi-scale information, enhancing the adaptability to fuzzy signals.
		\item Recognizing that traditional FC classifiers overfit easily and degrade performance with small datasets, we propose using an FCN classifier. This modification substantially improves generalization capabilities and maintains stable performance with limited data.
		\item We introduce a signal-guided spatial attention framework, enabling more effective feature weighting by focusing on critical signal components. When combined with transfer learning that leverages auxiliary datasets and the stable attention maps provided by ICVMD, our approach reduces the reliance on extensive labeled data and high-quality supplementary data, achieving at least 96\% accuracy in FS tests.
	\end{itemize}
	
	The remainder of this paper is structured as follows. Section \ref{sec2} details the generation of simulated signals to augment data diversity. Section \ref{sec3} outlines the architecture of our proposed method and explains the principles of each part. Section \ref{sec4} presents experiments on simulated and public datasets to compare the different methods. Then we analyze and summarize the results. Section \ref{sec5} concludes this paper.
	\section{Simulated Signal Generation and Description}\label{sec2}
	The signal types in actual datasets are usually limited, restricting the comprehensive evaluation of SEI methods \cite{comst}. We investigate the mechanisms underlying RFFs and generate some simulated signals with different IM types, enabling a detailed evaluation of the framework.
	\subsection{RFF Source Modeling} 
\begin{figure}[!t]
		\centering
		\includegraphics[width=0.9\linewidth]{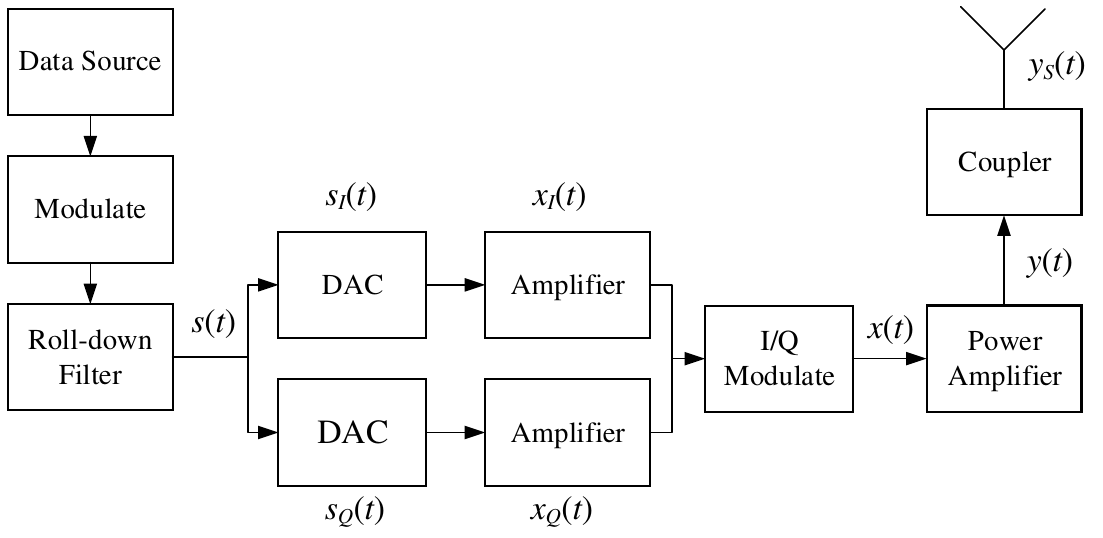}
		\caption{The signal generation and transmission process.}
		\label{fig2}
	\end{figure}

	UIMs originate from nearly any component within an emitter. To serve as reliable RFFs, these features must be universal, unique, and stable \cite{quzhongxin}. The most commonly used features include I/Q imbalance and power amplifier (PA) nonlinearity \cite{iqim1}. As illustrated in Fig.~\ref{fig2}, they are close to the output and considerably affect the signal. Multiple components influence the final I/Q imbalance, making it easily detectable but difficult to model uniformly for the purpose of reliable synthetic data generation \cite{iqim2}. Changes in the PA's physical properties are usually caused by gradual processes, such as temperature drift and aging, which take a long time to produce apparent discrepancies \cite{cycle}. This stability makes the PA a reliable source for RFFs, as the resulting features are difficult to replicate or forge in the short term \cite{imperf}.
	
	Since this study centers on the generation or reconstruction of stable and effective hardware fingerprints, a perfect PA functionality restoration is less critical. For a relatively simple device, we decompose it into components and reconstruct an equivalent simulation model by defining appropriate connections among them. Due to the complexity of actual systems, manual modeling becomes difficult; therefore, behavioral models offer a viable alternative to characterize PAs \cite{behavior}. They build the response connection between PA input and output by training data without knowing its internal structure. Considering the PA operates in the nonlinear region and historical inputs affect current outputs, we establish a nonlinear memory model \cite{nolin}.
	
	The Volterra model is widely used to model PAs with mildly nonlinear properties and high fidelity \cite{mod1}. However, it often results in a complex expression with many parameters to achieve a highly approximate simulation, which is slow to converge \cite{mod2}. While our study aims to achieve pre-distortion linearization, accurate PA modeling is a secondary concern. Thus, we choose the Hammerstein model \cite{ham-mod}, which meets performance requirements while keeping a simple structure. It comprises a memoryless nonlinearity followed by a linear time-invariant system, which can be written as
	\begin{equation}
		{y_H}\left( n \right) = \sum\limits_{q = 0}^{Q - 1} {{c_q}\sum\limits_{k = 1}^K {{b_k}x\left( {n - q} \right){{\left| {x\left( {n - q} \right)} \right|}^{k - 1}}} },
	\end{equation}
	where ${y_H}\left( n\right)$ is the output of the model, ${c_q}$ is the coefficient of the linear module with memory, and $b_k$ is the coefficient of the nonlinear module without memory. 
	\subsection{Simulation Parameters}
	\begin{table}
		\centering
		\caption{The nonlinear parameters of the emitters.}
		\label{tabsim}
		\begin{tabular}{@{}cccc@{}}
			\toprule
			Emitter & \begin{tabular}[c]{@{}c@{}}Nonlinear \\order $K$\end{tabular} & \begin{tabular}[c]{@{}c@{}}Memory \\depth $Q$ \end{tabular}&Model Coefficients $b_k$\\ \midrule
			Radiation1 & 5 & 6& {[}1,0,0.1126,0,0.2937{]} \\
			Radiation2 & 5 & 6& {[}1,0,0.2479,0,0.1396{]} \\
			Radiation3 & 5 & 6& {[}1,0,0.3959,0,0.1948{]} \\
			Radiation4 & 5 & 6& {[}1,0,0.5027,0,0.2833{]} \\
			Radiation5 & 5 & 6& {[}1,0,0.1683,0,0.4412{]} \\
			Radiation6 & 5 & 6& {[}1,0,0.3246,0,0.3463{]} \\
			Radiation7 & 5 & 6& {[}1,0,0.4698,0,0.3946{]} \\ \bottomrule
		\end{tabular}
	\end{table}
	Due to PA nonlinearity, variations in signal amplitude can cause substantial distortion, affecting the final performance evaluation. In the simulation, constant-envelope signals are utilized, and all signals are power-normalized. There are six typical signals with different parameters: constant wave (CW), linear frequency modulation (LFM), randomly coded binary phase shift keying (BPSK), quadrature PSK (QPSK), 8PSK and minimum shift keying (MSK). The parameters of seven emitters are listed in Table~\ref{tabsim}, the SNR range is set from -4dB to 20dB in steps of 2dB, the number of sampling points is 2100, and we generate 1500 training signals per transmitter. It should be noted that the primary objective of the simulation data is to provide a controlled environment for qualitative analysis and accurately assess the effectiveness of each component.
	\section{Proposed Method}\label{sec3}
	\subsection{Framework Overview}
	The implementation of our proposed ICVMD-SAT algorithm is illustrated in Fig.~\ref{fstr}. The framework consists of three components: an ICVMD module to decouple the IM and UIM components, a CNN-TCN backbone for robust sequence analysis, and a specialized branch network for signal-guided spatial attention. ICVMD provides a reference for the subsequent stages.The SAM uses the reconstructed signal, which reduces overfitting and dependency on auxiliary data in FS-SEI tasks by complementing the low-parameter FCN classifier. The specific operation and settings of each step are as follows:
	\begin{figure}[!t]
		\centering
		\includegraphics[width=0.7\linewidth]{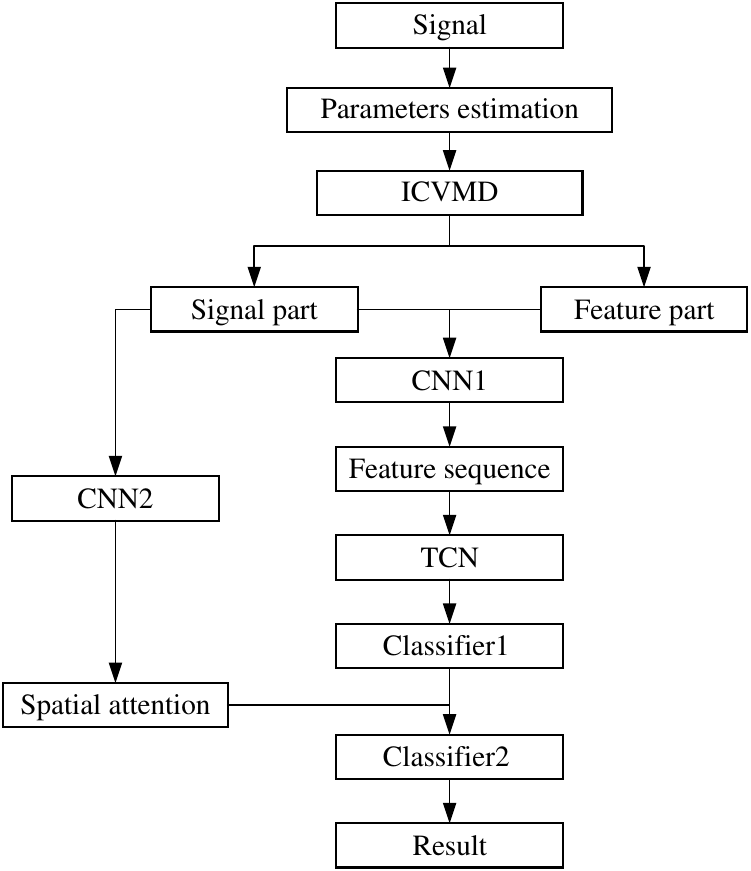}
		\caption{The structure of the ICVMD-SAT method.}
		\label{fstr}	
	\end{figure}
	
	(1) The first step is parameter estimation of the captured signal, which aims to collect enough information to set the ICVMD parameters, such as the range of $K$, the order of magnitude for $\alpha$, and the conditions each part requires. The goal is to derive stable and meaningful features.
	
	(2) By setting appropriate conditions, the modes are divided into signal and feature parts, and decisions are made on how to handle the DC component. The signal part is used to approximate the original signal and associate it with the feature part.
	
	(3) CNN1 is used to extract local features, and there are many network choices. The representative options include residual networks and complex value neural networks (CVNNs) \cite{CVNN}. We choose a standard residual architecture in experiments to ensure methodological generality. The resulting feature sequence is then processed by a TCN to model its long-range dependencies.
	
	(4) Classifier1 encodes each segment, outputting either a one-hot encoding of categories to implement a hard decision or feature logit vectors for a soft decision. Experiments indicated that the soft decision method yielded superior performance.
	
	(5) While CNN1's purpose is to extract fingerprint features, CNN2 analyzes the original signal using a SAM to generate attention weights. To ensure the output matches size, CNN2 uses a similar structure to CNN1 with fewer kernels to reduce the computational load, while ensuring that the final output matches the dimensions of CNN1. These attention weights are then used to element-wise weight the output from Classifier1, and the result is fed into a lightweight FCN Classifier2.
	\subsection{Variational Mode Decomposition}
	In fingerprint models, nonlinearity is mainly implemented by higher-order cumulants; thus, distinct features can be obtained by decomposing the signal. As our proposed ICVMD framework relies on this basic process, we first introduce the VMD algorithm. It fits signals iteratively without prior information to produce a series of signal components known as intrinsic mode functions (IMFs) and achieves non-recursive decomposition in both time and frequency domains, thereby highlighting features within the signal \cite{VMD}. These modes can reproduce specific attributes of the input signal to reconstruct it optimally.
	
	Affected by noise $\eta$, the received signal $f = {f_i} + \eta $, to recover the ideal signal $f_i$, we employ Tikhonov regularization:
	\begin{equation}
		\mathop {\min }\limits_{f_i} \left\{ {\left\| {f - {f_i}} \right\|_2^2 + \alpha \left\| {{\partial _t}{f_i}} \right\|_2^2} \right\},
	\end{equation}
	where the first term is a quadratic penalty term to determine the fidelity of the signal, and the second term makes the restored signal closer to the ideal one by using the square of gradient. We set the number of modes $K$ and initialize all modes, thereby calculating the one-sided analytic spectra through the Hilbert transform and shift them to baseband. This formulation results in a constrained variable problem:
	\begin{equation}
		\begin{array}{*{20}{r}}
			{\mathop {\min }\limits_{\left\{ {{u_k}} \right\},\left\{ {{\omega _k}} \right\}} \left\{ {\sum\limits_k {\left\| {{\partial _t}\left[ {\left( {\delta (t) + \frac{j}{{\pi t}}} \right) * {u_k}(t)} \right]{e^{ - j{\omega _k}t}}} \right\|} _2^2} \right\}}\\
			{{\rm{\;s}}{\rm{.t}}{\rm{.}}\mathop \sum \limits_k {u_k} = f},
		\end{array}
	\end{equation}
	where $u_k$ and $\omega_k$ represent the mode and its center frequency. We use Lagrange multipliers $\lambda$ to make the problem unconstrained. The problem is transformed into:
	
	\begin{align}\label{eq5}
		{\cal L}&\resizebox{0.95\hsize}{!}{$\left( {\left\{ {{u_k}} \right\} ,\left\{ {{\omega _k}} \right\},\lambda } \right): = \alpha \sum\limits_k {\left\| {{\partial _t}\left[ {\left( {\delta (t) + \frac{j}{{\pi t}}} \right) * {u_k}(t)} \right]{e^{ - j{\omega _k}t}}} \right\|_2^2}$} \notag \\
		&\resizebox{0.8\hsize}{!}{$\quad + \left\| {f(t) - \mathop \sum \limits_k {u_k}(t)} \right\|_2^2 + \left\langle {\lambda (t),f(t) - \mathop \sum \limits_k {u_k}(t)} \right\rangle.$}
	\end{align}
	
	In Eq. \ref{eq5}, the first term minimizes the cumulative bandwidth to ensure spectral compactness and prevent mode mixing, with the factor $\alpha$ balancing the regularization strength against reconstruction fidelity. The quadratic penalty and $\lambda$ jointly enforce the strict reconstruction constraint, where $\lambda$ iteratively adjusts mode deviations to ensure the decomposition is numerically precise. The $u_k$ and $\omega_k$ are updated in the Fourier domain by the Alternating Direction Method of Multipliers (ADMM) algorithm \cite{admm}, respectively:
	\begin{equation}
		\label{eq-u}
		u_k^{n + 1}(\omega ) = \frac{{f(\omega ) - \sum\limits_{i \ne k} {{u_i}(\omega ) + \frac{{\lambda (\omega )}}{2}} }}{{1 + 2\alpha {{(\omega- {\omega _k})}^2}}}.
	\end{equation}
	\begin{align}
		\omega _k^{n + 1} = \frac{{\int_0^\infty{\omega {{\left| {{u_k}(\omega )} \right|}^2}d\omega } }}{{\int_0^\infty{{{\left| {{u_k}(\omega )} \right|}^2}d\omega } }}.\label{eq-w}
	\end{align}
	
	We can find $\omega_k$ corresponds to the gravity center of the mode's power spectrum, and the complete process is summarized as Algorithm.~\ref{alg1}.
	\begin{algorithm}
		\renewcommand{\algorithmicrequire}{\textbf{Initialize}}
		\renewcommand{\algorithmicensure}{\textbf{Output:}}
		\caption{Complete optimization process of VMD}
		\label{alg1}
		\begin{algorithmic}[1]
			\REQUIRE{$\{ \hat{u}_k^1\} ,\{\hat{\omega} _k^1\} ,{\hat{\lambda} ^1},n \leftarrow 0$ } 
			\REPEAT 
			\STATE {$n \leftarrow n+1$}
			\FOR{$k=1$ to $K$}
			\STATE Update $\hat{u}_k$ for all $\omega \ge 0$ by Eq.~\ref{eq-u}: 			
			\STATE Update $\hat{\omega}_k$ by Eq.~\ref{eq-w}: 
			\ENDFOR
			\STATE Dual ascent for all $\omega \ge 0$:		
			$${\hat \lambda ^{n + 1}}(\omega ) \leftarrow {\hat \lambda ^n}(\omega ) + \tau \left( {\hat{f}(\omega ) - \sum\limits_k {\hat{u}_k^{n + 1}(\omega )} } \right)$$
			\UNTIL{convergence: $\sum\nolimits_k {{{\left\| {\hat u_k^{n + 1} - \hat u_k^n} \right\|_2^2} \mathord{\left/
							{\vphantom {{\left\| {\hat u_k^{n + 1} - \hat u_k^n} \right\|_2^2} {\left\| {\hat u_k^n} \right\|_2^2}}} \right.
							\kern-\nulldelimiterspace} {\left\| {\hat u_k^n} \right\|_2^2}}}< \epsilon $}			
		\end{algorithmic}
	\end{algorithm}
	\subsection{Integrated Complex Variational Mode Decomposition}
	Although decomposing the input signal enhances recognition, the standard VMD algorithm has several limitations for SEI tasks. Each mode obtained in existing studies lacks precise meanings, making it insufficient for fully utilizing information. VMD is defined only for real-valued signals; thus, the rising adoption of complex-valued signals drives an urgent demand for extending it to the complex domain, and motivates our proposal of the ICVMD.
	
	There are interdependencies between the real and imaginary parts of a complex-valued signal, which results in information loss when applying VMD separately to the two parts and adding them accordingly \cite{CVMD}. Moreover, the modes obtained from the two parts are not guaranteed to match perfectly due to I/Q imbalance. In the ICVMD method, we decompose the signal and separate the positive and negative frequency parts by a step function $H(\cdot)$, then we obtain their corresponding real-valued signals, respectively:
	\begin{gather}
		x_+(t)=\Re\left\{\mathcal{F}^{-1}(H(e^{j\omega})X(e^{j\omega}))\right\}
		\\
		x_-(t)=\Re\left\{\mathcal{F}^{-1}(H(e^{j\omega})X(e^{-j\omega}))\right\},
	\end{gather}
	where $\mathcal{F}^{-1}$ denotes the inverse Fourier transform, and $\Re$ represents the operator to extract real part. The signals by VMD can be expressed as Eq.~\ref{eq10} and \ref{eq11}:
	\begin{gather}
		\label{eq10}
		{x_ + }[n] = \sum\limits_{i = 1}^{{N_ + }} {{x_i}[n] + {r_ + }[n]} 
		\\\label{eq11}
		{x_ - }[n] = \sum\limits_{i =- {N_ - }}^{ - 1} {{x_i}[n] + {r_ - }[n]}. 
	\end{gather}
	The final signal can be represented as Eq.~\ref{fie} by Hilbert transform $\mathcal{H}(\cdot)$.
	\begin{equation}
		x[n] = ({x_ + }[n] + j{\mathcal H}[{x_ + }[n]]) + {({x_ - }[n] + j{\mathcal H}[{x_ - }[n]])^*} \label{fie}
	\end{equation}
	
	The parameter settings in VMD, particularly $K$ and $\alpha$, affect the quality of extracted information. In general, the search for suitable parameter range relies on experience and basic knowledge regarding the signal type and distribution. To mitigate this dependency, the particle swarm optimization (PSO) algorithm \cite{vmds} is used to obtain the initial optimal parameter pair ($K \in [5, 20], \alpha \in [500, 5000]$). Since the entropy component introduces a negative sign, maximizing the mean log energy-entropy fitness is equivalent to minimizing the following objective:
	\begin{equation}
		\left\langle {\alpha ,K} \right\rangle = \arg \mathop {\min }\limits_{\{ \alpha ,K\} } \left\{ {\frac{1}{K}\sum\limits_{k = 1}^K {\log {{\left( {{R_k}(t)} \right)}^2}} } \right\},
	\end{equation}
	where ${R_k}(t)$ denotes the $k$-th normalized IMF, but the  optimal values depend on the specific signal rather than the type. The results of VMD on similar signals can vary obviously even under identical settings, leading to poor stability in non-cooperative scenarios. When signals are input into an NN for unified processing, this lack of stability becomes a critical bottleneck, as NN performance often prioritizes consistent feature representation over the separation effect of individual modes.
	
	The ICVMD combines independent modes to represent different signal parts according to the task and scenario, giving them the corresponding meanings. The signal is typically aggregated into a signal part (comprising high-energy modes consistent across the protocol) and a feature part (capturing high-variance residuals and parasitic distortions). This partitioning establishes a stable reference for IM, enabling the network to isolate UIM as hardware-specific interactions. The direct current (DC) component and special features strongly related to the transmitter are considered independently if necessary. In contrast, the CVMD proposed in \cite{CVMD} is merely a generic complex-domain extension of VMD. It shares parameters across positive and negative frequency domains, which is a simplification unsuitable for many signals, and it neglects to utilize the DC component. Since it relies on VMD for implementation, the parameter selection challenge persists. While the ICVMD process is affected by $K$, it exhibits superior distributional robustness. When $K$ varies within a small range, results become stable because it typically causes some modes to split or reorganize, which does not significantly affect the subsequent modes recombination. Due to the introduction of the screening mechanism in ICVMD, more modes are used for accurate feature extraction (typically ranging from 10 to 15, compared to the standard 5 to 8). The ICVMD provides greater flexibility in the SEI field by reducing reliance on prior information.
	\subsection{Temporal Convolutional Network}
	We apply TCN to analyze the feature sequences extracted from the ICVMD signals. Traditional RNNs are suited for processing maturely encoded discrete sequences, but they are less effective in classification tasks compared to CNNs \cite{seq-com}. TCN aligns with the requirements for sequence analysis in SEI by combining feature extraction with sequence modeling, and the operations can be performed in parallel \cite{TCN}. Its operational principles are detailed below:
	
	\begin{figure}[!t]
		\centering
		\includegraphics[width=0.85\linewidth]{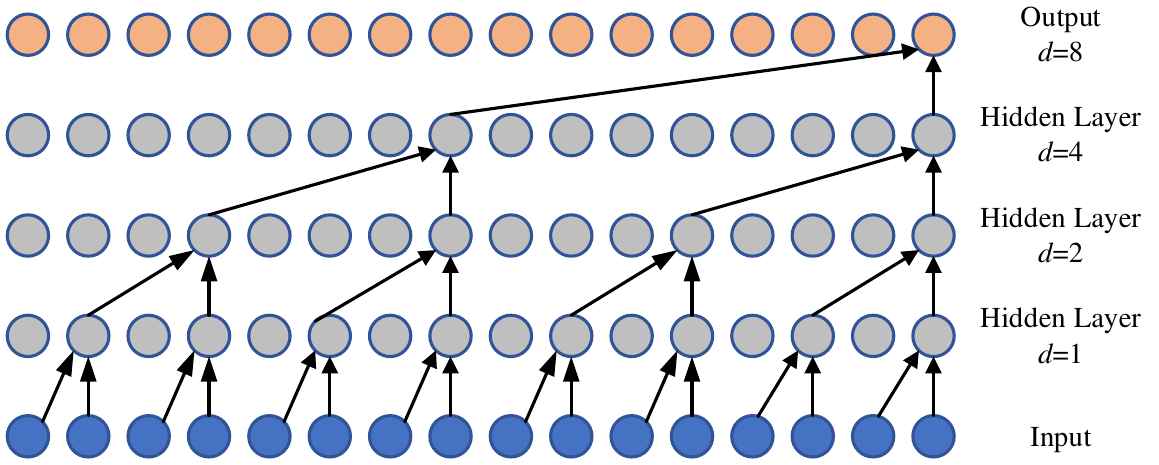}
		\caption{Causal dilation convolution structure ($d={1,2,4,8}$, $l=2$).}
		\label{dilatp2}
	\end{figure}
	To capture long-term temporal dependencies while preventing future information leakage, TCN utilizes causal dilated convolutions. It allows kernels to skip input at a fixed stride by interspersing $k-1$ zeros between two adjacent taps, and $k$ is called the dilation rate \cite{dilat}. It increases the feature map resolution while not reducing the receptive field of every neuron, ensuring the model performs stably in long sequence tasks. The causal dilated convolution operation at the $i$th layer can be represented as:
	\begin{equation}\label{eq13}
		v_t^{(i + 1)} = \sum\limits_{j = 0}^{l - 1} {g_j^{(i)}v_{t - (l - 1 - j)d^{(i)}}^{(i)}}.
	\end{equation}
	where $d^{i}=k^{i-1}$ denotes the expansion factor of the layer, and $d$ increases exponentially with $i$ from Fig.~\ref{dilatp2}, so we can add $k$ to expand the receptive field. The left side of the input sequence is padded with $l-1$ zeros, and $l$ is the convolution kernel size, which keeps the output length equal to the input and makes them correspond to each other in position.
	
	Meanwhile, TCN utilizes residual connections to enhance network memory and facilitate information sharing across layers. A residual block defines the output as $o = x + F(x)$, where $x$ is the input and $F$ represents the network transformation. Training the residual $F(x)$ instead of the entire transformation allows the model to integrate outputs from each layer, leading to faster and more stable convergence.
	
\begin{figure}[!t]
		\centering
		\includegraphics[width=0.6\linewidth]{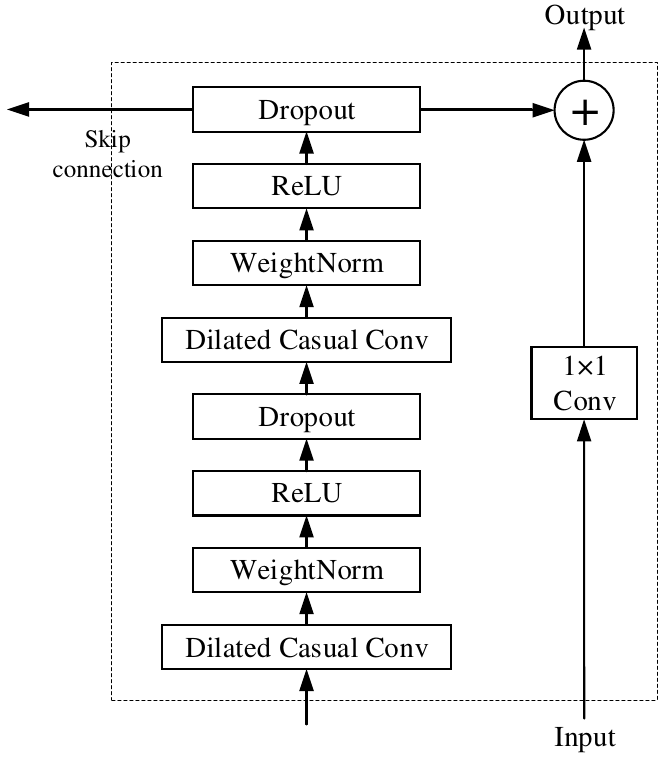}
		\caption{The structure of a simple TCN block.}
		\label{tcn-blo}
	\end{figure}

	The basic block of TCN is demonstrated in Fig.~\ref{tcn-blo}, and it has been proven in \cite{my} that applying TCN in signal classification tasks like SEI increases training speed and provides stable performance. To further enhance the model's structural flexibility against unpredictable signal variations, we construct a multi-scale TCN by adjusting the dilation factors across different layers. While a larger receptive field captures broader contextual information, a smaller range focuses on localized details. These convolutional layers act as adaptive band-pass filters, which can extract embedded signal features from various frequency bands to obtain comprehensive information. We stack the outputs of different dilated convolutional layers and apply a $1 \times 1$ convolution for feature fusion. This multi-scale approach enables the network to receive information across different temporal scales, enhancing its adaptability to complex actual signals.
	\subsection{Spatial Attention Mechanism}
	FC classifiers, which contain numerous parameters and require a fixed-size input, are prone to overfitting with small sample sizes. Furthermore, their reliance on a flattened input makes them sensitive to feature order, a factor that is irrelevant in our task, so we constructed an FCN to address this issue \cite{fcn}. Specifically, the feature map output by the TCN backbone is first processed by a global pooling layer, which reduces the input sequence to a single vector while preserving channel-wise semantics. This vector is then fed into a final $1 \times 1$ convolutional layer with $N_{class}$ filters. This FCN design improves the parameter efficiency, providing a structural regularization effect that facilitates stable generalization. However, the simple FCN is suboptimal, as it treats features of each segment equally or ignores some, while some are more significant or less affected by noise. When the signal recognition is biased, there may be purely noise segments in the steady-state signal that affect the recognition, necessitating the redistribution of the feature weights.
	
	Attention mechanisms enable the network concentrate on prominent parts within the feature maps, reducing computation and improving accuracy. Eq.~\ref{att} can summarize the principle:
	\begin{equation}
		{\rm{Attention = }}f\left( {g(x),x} \right),
		\label{att}
	\end{equation}
	where $g(x)$ represents generating attention to distinguish information from different regions, and $f(g(x),x)$ denotes processing input $x$ based on $g(x)$. Typical self-attention calculates weights via query-key similarities; however, such purely data-driven weighting often lacks physical interpretability in RFF extraction.
	
	SAMs usually model attention as spatial probabilities to reweight different positions within the feature map, aiming to select spatial regions adaptively. Some research integrates attention directly into hidden layers to form a block \cite{cbam}. It can be used in any CNN architecture with minimal computational overhead. The block generally requires sufficient data to generate attention, and the model is highly targeted with the cooperation of different layers, which is detrimental to attention transfer in our purpose.
	
	\begin{figure}[!t]
		\centering
		\includegraphics[width=0.75\linewidth]{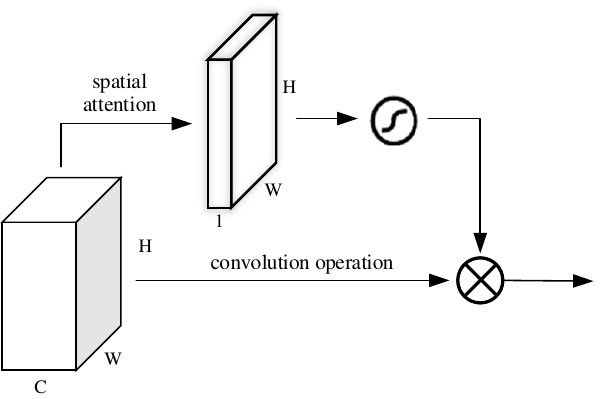}
		\caption{A schematic diagram of the spatial attention mechanism.}
		\label{sam}
	\end{figure}
	As shown in Fig.~\ref{fstr} and Fig.~\ref{sam}, our framework introduces a parallel branch network that receives IM input. The branch focuses solely on the feature importance of signal segments to determine attention and affect the backbone network output \cite{sam1}. This design uses the IM signal to generate attention weights, providing a consistent reference across different devices using the same protocol. By aligning the attention mechanism directly with the IM state, the model adapts to domain shifts without requiring additional alignment losses. Transformation-based methods like the spatial transformer network \cite{stn} can achieve spatial invariance to simplify inference. However, we do not adopt them due to non-linear relation between UIMs/IMs in spatial transforms and the high risk of information loss. We apply the spatial attention transfer (SAT) on the branch, which trains models on similar tasks and provides knowledge of which areas should be focused on because the attention relying on the shared IM features is transferable.
	
	Whether restoring the original signal by ICVMD or executing SAT, the process relies on the IM type in our method. This IM result remains independent of sample size since reliable parameter settings can be acquired. Although RFFs are difficult to imitate, the corresponding IM signals are easy to generate. Similar signals can even be emitted directly from the device by data augmentation or extracting the parameters and types, thus suitable extra data for parameter acquisition is available. It is only necessary to ensure that the new transmitter conforms to the same modeling framework without requiring the specific hardware parameters. This maintains consistent importance weighting across different hardware entities.
	
	\section{Simulation and Analysis}\label{sec4}
	For all experiments, we split a portion of each dataset for validation based on its size. The models are implemented in PyTorch and trained on NVIDIA GeForce RTX 2070 and 3080 GPUs. We employ the Adam optimizer with $\beta_1 = 0.9$, $\beta_2 = 0.999$, and a batch size of 128. The initial learning rate is set to $1 \times 10^{-3}$, which follows a step decay to $1 \times 10^{-6}$. In the transfer learning phase, we first freeze the branch attention network to train the backbone and FCN modules, and then perform a global fine-tuning stage. To ensure stable convergence, an early stopping strategy with a patience of 5 epochs is applied based on the validation loss. Additionally, a consistent set of random seeds $\{2, 3, 4, \cdots\}$ is utilized across trials to ensure statistical reliability.
	\subsection{Signal Representation Functional Verification}
	Before evaluating the performance on SEI tasks, it is essential to verify the quality of signal representation and reconstruction achieved by ICVMD, which is a prerequisite for our method. We conducted experiments on the RML2016.10b dataset, which is part of a public modulation recognition dataset repository provided by the DeepSig team \cite{rml1}. The dataset includes ten types of signals: 8PSK, AM-DSB, BPSK, CPFSK, GFSK, PAM4, QAM16, QAM64, QPSK, and WBFM, with SNR ranging from -18dB to 18dB, in 2dB increments.
	
	\begin{figure}[!t]
		\centering
		\includegraphics[width=0.85\linewidth]{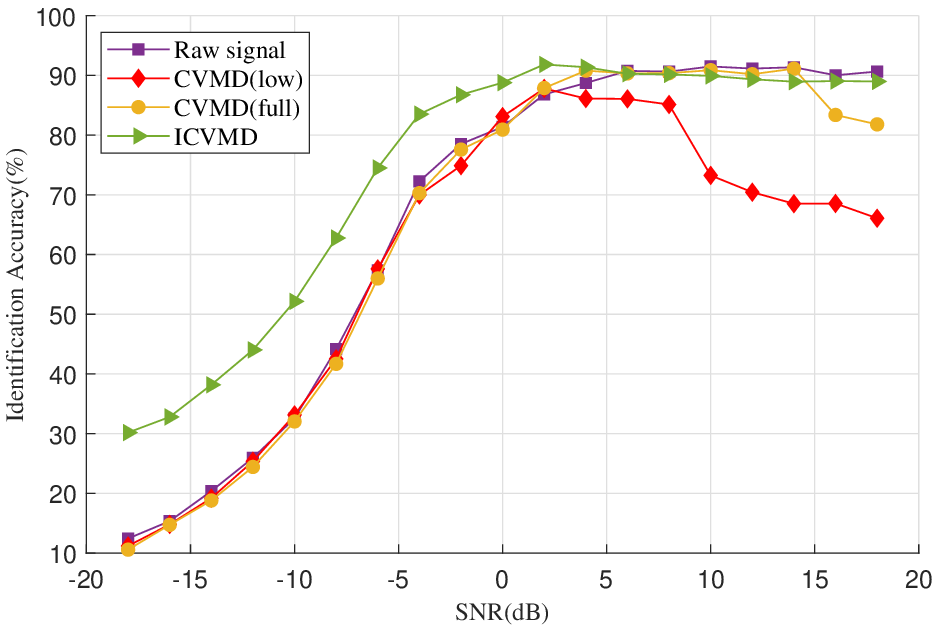}
		\caption{The modulation recognition accuracy in the RML dataset.}
		\label{modrec}
	\end{figure}
	An official benchmark classifier was employed for evaluation, as optimizing the classifier itself was not our objective. The input included the raw signals, the low-order IMFs from CVMD (CVMD-low), all modes obtained by CVMD (CVMD-full), and the restored signals using ICVMD. Fig.~\ref{modrec} shows that the ICVMD signals are less affected as the SNR decreases, outperforming other methods and validating its restoration capability at low SNR. The mode number $K$ is fixed across all SNR levels for experimental consistency, and the signals in the dataset are split into short segments. At high SNR, the lack of noise to accommodate residual components leads to mode splitting, causing instability in the feature representation and a slight performance degradation in these methods. In this real-world dataset, high-fidelity noise transients at high SNR may lead to inadvertent overfitting, whereas lower SNR provides a form of natural regularization.
	
	\begin{figure}
		\centering
		\subfloat[]{		
			\includegraphics[width=0.65\linewidth]{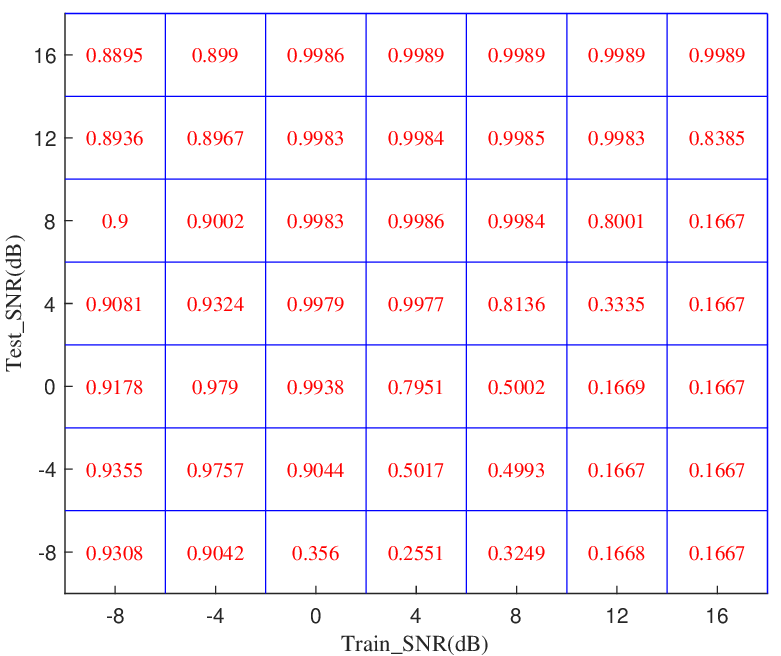}
			\label{suba}
		}
		\\
		\subfloat[]{		
			\includegraphics[width=0.65\linewidth]{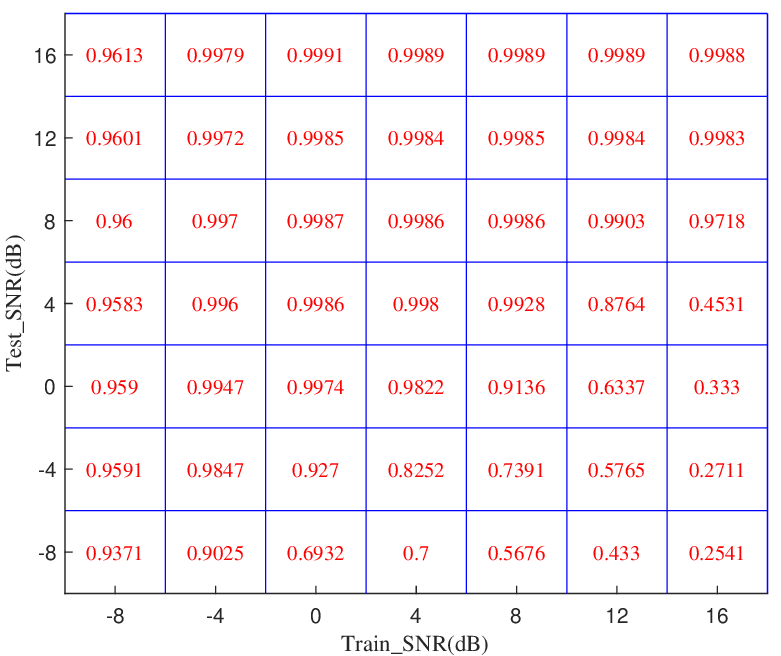}
			\label{subb}
		}
		\caption{Simulated modulation cross-recognition accuracy under SNR mismatch. (a) Original signal. (b) ICVMD reconstructed signal.}
		\label{modvar}
	\end{figure}
	To further probe the robustness of the ICVMD method under mismatched conditions, we compared cross-recognition performance under different SNRs. The experiment was carried out using the simulated data to avoid the effect of SNR on RML results. As shown in Fig.~\ref{modvar}, classification accuracy generally decreases when the test SNR is lower than that of the training set. Comparing Fig.~\ref{modvar}\subref{suba} and Fig.~\ref{modvar}\subref{subb}, the recognition accuracy with ICVMD generally surpasses that of the original signals. The difference is apparent when using a high SNR model to recognize low SNR signals. With training and test SNRs at 16dB and 8dB, respectively, it is difficult for the original input model to recognize it correctly, whereas the accuracy remains at 97\% after using ICVMD. This further verifies ICVMD's ability to recover the underlying signal and suppress noise.
	\subsection{Simulated Data}
	\begin{figure*}
		\centering
		\subfloat[]{
			\includegraphics[width=0.4\linewidth]{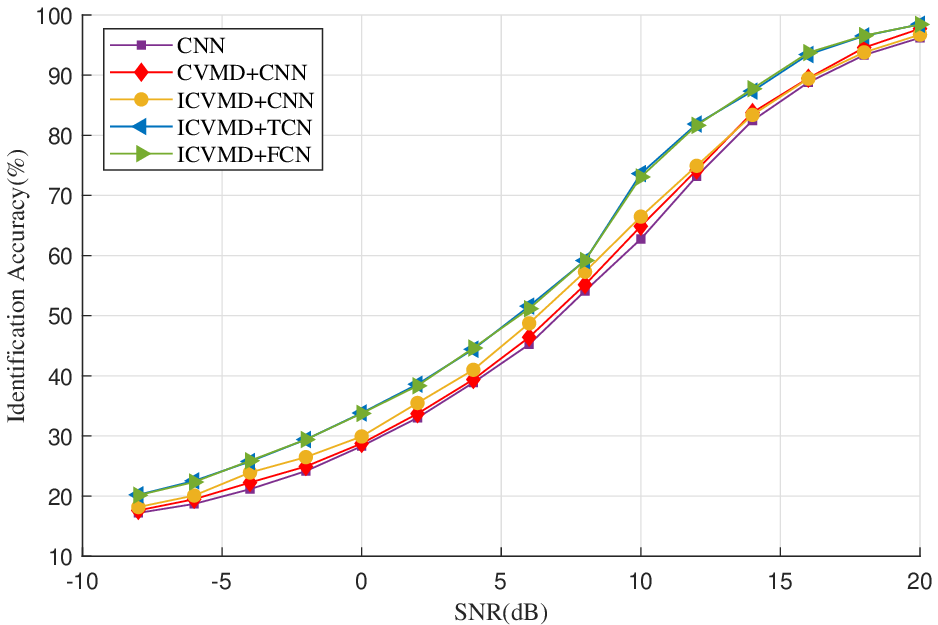}
			\label{xya}
		}
		\hfil
		\subfloat[]{
			\includegraphics[width=0.4\linewidth]{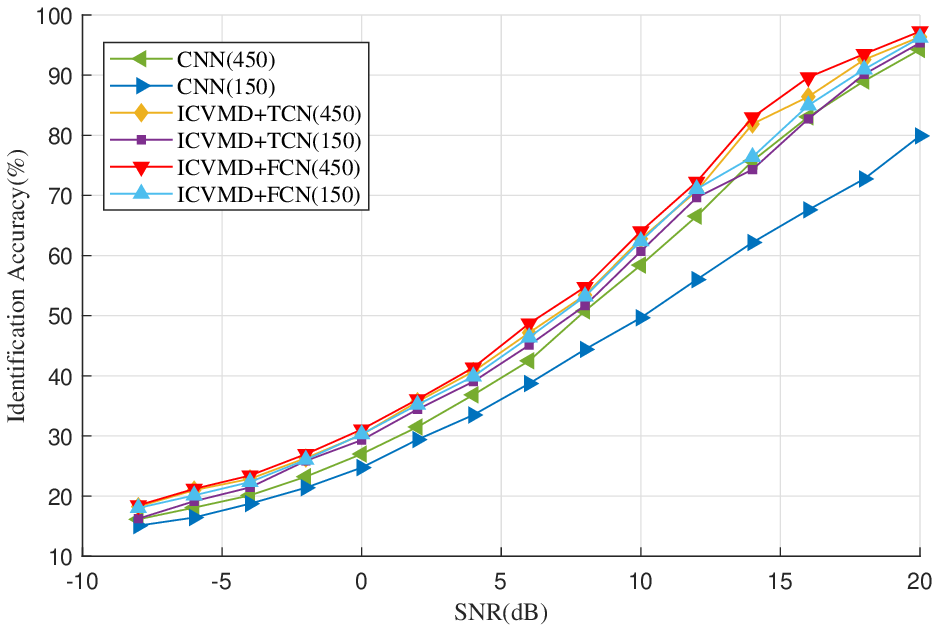}
			\label{xyb}
		}
		\caption{Experimental results on the simulated dataset. (a) Ablation experiment using all data. (b) Ablation experiment using small samples.}
		\label{picfull}
	\end{figure*}
	\begin{table*}
		\centering
		\caption{Recognition result after transfer learning}
		\label{tabqy}
		\begin{tabular}{@{}cccccccccccc@{}}
			\toprule
			Accuracy (\%) & \begin{tabular}[c]{@{}c@{}}Sample\\ proportion\end{tabular} & 0dB & 2dB & 4dB & 6dB & 8dB & 10dB & 12dB & 14dB & 16dB & 18dB \\ \midrule
			Best & 100\% & 34.1 & 38.3 & 45.4 & 52.2 & 60.1 & 74.3 & 82.5 & 88.9 & 94.2 & 96.9 \\ \midrule
			\multirow{3}{*}{Proposed} & 30\% & 32.6 & 38.2 & 43.7 & 51.1 & 59.4 & 72.6 & 82.0 & 88.5 & 93.6 & 96.7 \\
			& 10\% & 32.1 & 37.8 & 43.4 & 50.2 & 57.6 & 72.2 & 81.1 & 87.0 & 92.8 & 95.9 \\
			& 3\% & 31.5 & 37.1 & 42.6 & 49.5 & 56.1 & 70.8 & 79.4 & 85.0 & 91.2 & 93.8 \\ \midrule
			\multirow{3}{*}{w/o FCN} & 30\% & 32.3 & 38.0 & 43.5 & 49.7 & 56.1 & 70.5 & 78.3 & 85.3 & 93.0 & 94.7 \\
			& 10\% & 31.6 & 37.2 & 42.6 & 48.5 & 55.1 & 69.2 & 77.3 & 82.2 & 90.4 & 91.5 \\
			& 3\% & 31.2 & 36.4 & 41.3 & 47.0 & 52.4 & 67.3 & 76.2 & 81.0 & 87.7 & 88.9 \\ \midrule
			\multirow{3}{*}{w/o TCN} & 30\% & \multicolumn{1}{l}{32.0} & \multicolumn{1}{l}{37.6} & \multicolumn{1}{l}{43.2} & \multicolumn{1}{l}{49.1} & \multicolumn{1}{l}{54.2} & \multicolumn{1}{l}{67.8} & \multicolumn{1}{l}{77.2} & \multicolumn{1}{l}{83.2} & \multicolumn{1}{l}{91.5} & \multicolumn{1}{l}{94.5} \\
			& 10\% & \multicolumn{1}{l}{31.2} & \multicolumn{1}{l}{36.7} & \multicolumn{1}{l}{41.9} & \multicolumn{1}{l}{48.0} & \multicolumn{1}{l}{53.5} & \multicolumn{1}{l}{65.5} & \multicolumn{1}{l}{74.8} & \multicolumn{1}{l}{81.2} & \multicolumn{1}{l}{87.6} & \multicolumn{1}{l}{89.4} \\
			& 3\% & \multicolumn{1}{l}{30.8} & \multicolumn{1}{l}{35.9} & \multicolumn{1}{l}{40.4} & \multicolumn{1}{l}{46.7} & \multicolumn{1}{l}{52.0} & \multicolumn{1}{l}{64.1} & \multicolumn{1}{l}{72.5} & \multicolumn{1}{l}{78.6} & \multicolumn{1}{l}{83.4} & \multicolumn{1}{l}{86.6} \\ \midrule
			\multirow{3}{*}{w/o ICVMD} & 30\% & \multicolumn{1}{l}{31.1} & \multicolumn{1}{l}{36.9} & \multicolumn{1}{l}{42.2} & \multicolumn{1}{l}{48.7} & \multicolumn{1}{l}{52.7} & \multicolumn{1}{l}{66.8} & \multicolumn{1}{l}{75.9} & \multicolumn{1}{l}{81.5} & \multicolumn{1}{l}{89.8} & \multicolumn{1}{l}{94.1} \\
			& 10\% & \multicolumn{1}{l}{30.3} & \multicolumn{1}{l}{35.0} & \multicolumn{1}{l}{40.1} & \multicolumn{1}{l}{46.3} & \multicolumn{1}{l}{51.4} & \multicolumn{1}{l}{63.9} & \multicolumn{1}{l}{72.4} & \multicolumn{1}{l}{79.8} & \multicolumn{1}{l}{84.9} & \multicolumn{1}{l}{88.5} \\
			& 3\% & \multicolumn{1}{l}{29.4} & \multicolumn{1}{l}{33.3} & \multicolumn{1}{l}{38.7} & \multicolumn{1}{l}{44.6} & \multicolumn{1}{l}{50.3} & \multicolumn{1}{l}{61.7} & \multicolumn{1}{l}{69.7} & \multicolumn{1}{l}{76.1} & \multicolumn{1}{l}{80.7} & \multicolumn{1}{l}{84.4} \\ \midrule
			\multirow{3}{*}{\begin{tabular}[c]{@{}c@{}}CNN\\ (max)\end{tabular}} & 30\% & 32.5 & 37.3 & 42.0 & 48.6 & 59.5 & 70.4 & 78.4 & 85.7 & 91.5 & 95.0 \\
			& 10\% & 30.7 & 35.4 & 40.8 & 47.3 & 54.3 & 66.5 & 76.9 & 84.3 & 90.3 & 94.8 \\
			& 3\% & 30.1 & 33.8 & 39.7 & 46.2 & 51.4 & 66.1 & 74.9 & 82.9 & 88.7 & 93.5 \\ \midrule
			\multirow{3}{*}{\begin{tabular}[c]{@{}c@{}}CNN\\ (min)\end{tabular}} & 30\% & 31.6 & 34.8 & 36.3 & 44.0 & 54.8 & 61.3 & 71.1 & 75.2 & 79.8 & 89.7 \\
			& 10\% & 30.6 & 32.8 & 37.1 & 43.1 & 51.1 & 54.5 & 60.0 & 64.1 & 73.3 & 78.1 \\
			& 3\% & 28.5 & 31.2 & 35.1 & 41.2 & 49.2 & 53.3 & 58.6 & 63.7 & 68.7 & 73.9 \\ \bottomrule
		\end{tabular}
	\end{table*}
	We conducted an ablation experiment using simulated data to verify the effectiveness of each component, and the results are presented in Fig.~\ref{xya}. The CNN and TCN models were designed with similar structures to keep comparable complexity. The performance consistently improved as the method transitioned from conventional approaches to ours (starting from a baseline CNN, we incrementally introduced CVMD, replaced it with ICVMD, adopted the TCN module, and finally switched to the FCN classifier). When using the same CNN model, the CVMD-based method performed better under high SNRs. Decreasing SNR leads to variations in optimal parameters for signals, causing the CVMD output to begin destabilizing and affect recognition; nevertheless, this approach still outperforms direct classification of original signals. Once the TCN backbone is integrated, the recognition accuracy improves and remains stable. Incorporating the spatial attention module yielded the best results on the complete dataset. Although it did not notably enhance the absolute accuracy, the classification became more stable, with accuracy consistently exceeding the previous mean. The TCN model with an FC classifier outperformed the FCN method with the curves nearly overlapping when using all data, but the opposite result when the data decreased, as shown in Fig.~\ref{xyb}.
	
	We generated data for transfer training, and the five new emitters parameters differed entirely from the training set. The results in Table~\ref{tabqy} indicate that all the methods perform better after introducing models on similar tasks. Previous studies often compared proposed methods against traditional ones without granting the latter access to extra data, leading to inconsistent conditions and unfair comparisons. In our experiments, applying transfer learning to the traditional method improved its performance, with its lower accuracy bound often exceeding its previous mean value. However, the stability of the traditional CNN remained inferior to the proposed framework and its ablation variants. Even when utilizing only 3\% of the target data, the performance of our method decayed by less than 5\% compared to the optimal results, demonstrating its effectiveness in FS scenarios. Ablation studies were conducted to verify the contribution of each component. The results show that removing any module leads to a performance decline. Excluding the FCN had the least impact on the final accuracy. In contrast, the most significant performance degradation occurred when ICVMD was replaced by direct raw signal input. All ablation versions consistently outperformed the enhanced traditional CNN, validating the effectiveness of our proposed architecture.
	\subsection{Actual Dataset and Baselines}
	To validate our method in real-world scenarios, we conducted experiments on actual data from a public dataset \cite{cycle}, which has 60 ZigBee devices. All ZigBee modules operate at a 2505MHz carrier frequency and a maximum transmission power of 8dBm. Some of them are equipped with RFX2401C chips to increase the transmit power, with a saturation power of 22dBm. The data in ZigBee frames is random to simulate real-world scenarios, and the receiver's sampling rate was configured to 10MHz. The environments are indoors and outdoors with LOS and non-line-of-sight (NLOS) propagation.
	
	\begin{figure*}
		\centering
		\subfloat[]{		
			\includegraphics[width=0.24\linewidth]{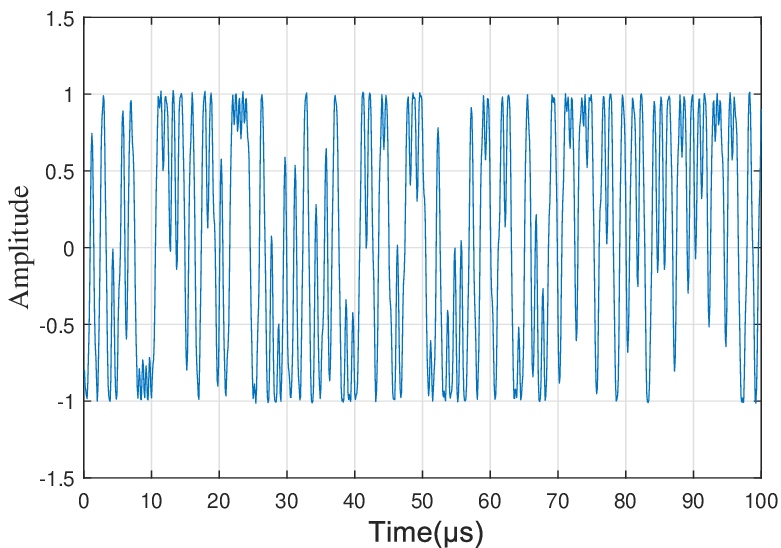}
		}
		\subfloat[]{		
			\includegraphics[width=0.24\linewidth]{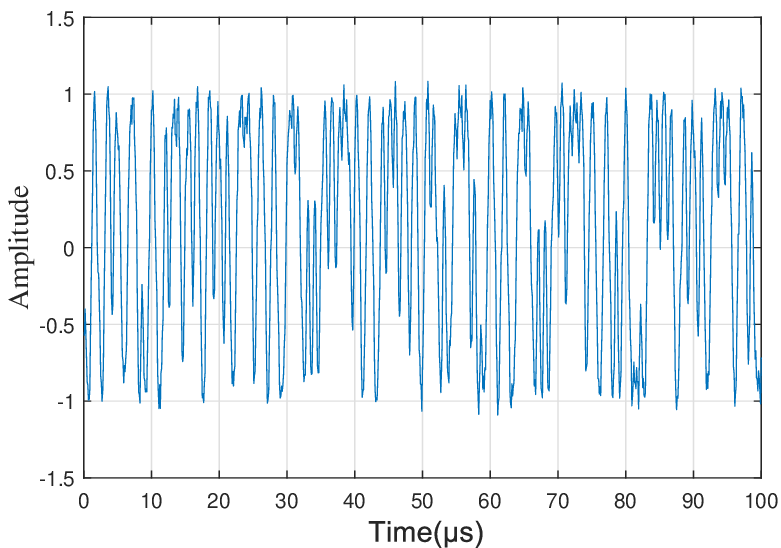}
		}
		\subfloat[]{	
			\includegraphics[width=0.24\linewidth]{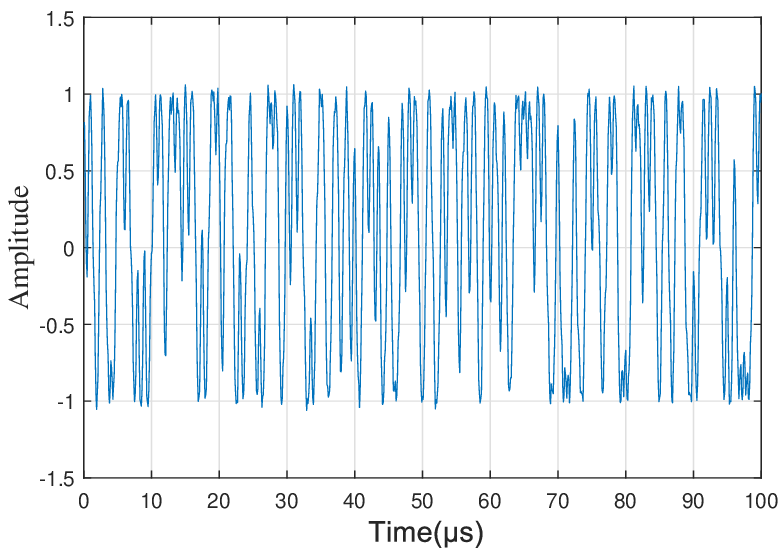}
		}
		\subfloat[]{		
			\includegraphics[width=0.24\linewidth]{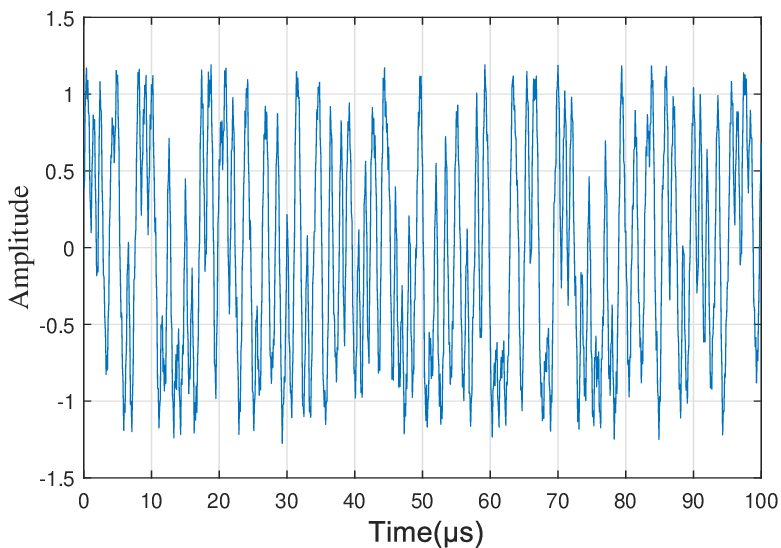}
		}
		\caption{Zigbee signal time domain graphs. (a) Indoor-LOS. (b) Indoor-NLOS. (c) Outdoor-LOS. (d) Outdoor-NLOS.}
		\label{sigp51}
	\end{figure*}
	\begin{figure*}
		\centering
		\subfloat[]{		
			\includegraphics[width=0.24\linewidth]{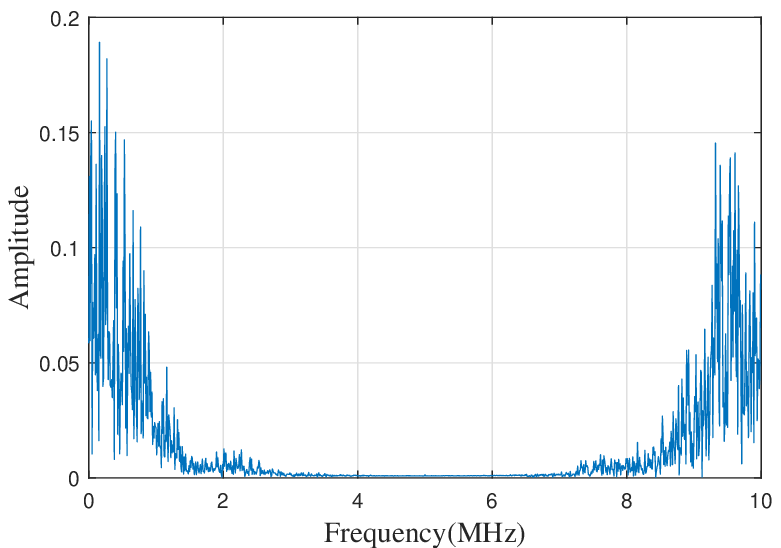}
		}
		\subfloat[]{		
			\includegraphics[width=0.24\linewidth]{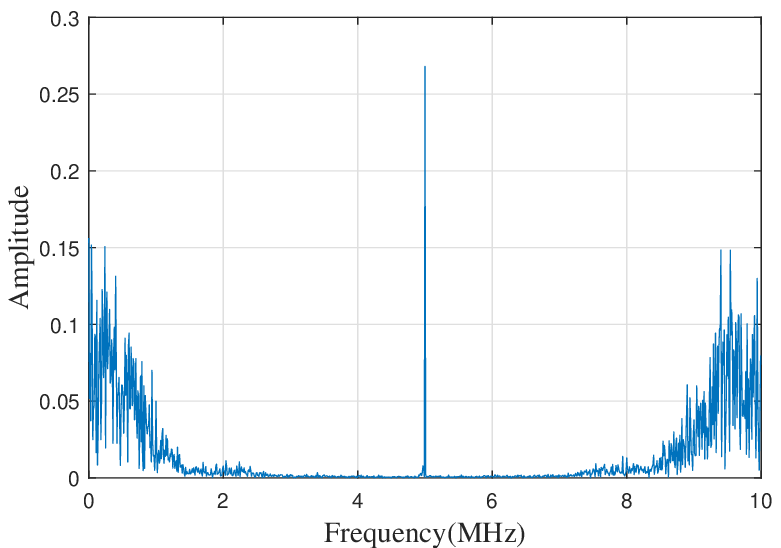}
		}	
		\subfloat[]{	
			\includegraphics[width=0.24\linewidth]{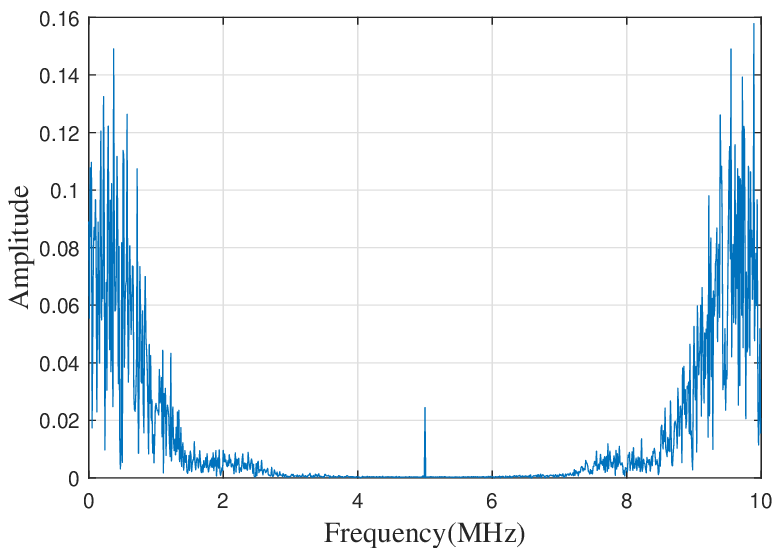}
		}
		\subfloat[]{
			\includegraphics[width=0.24\linewidth]{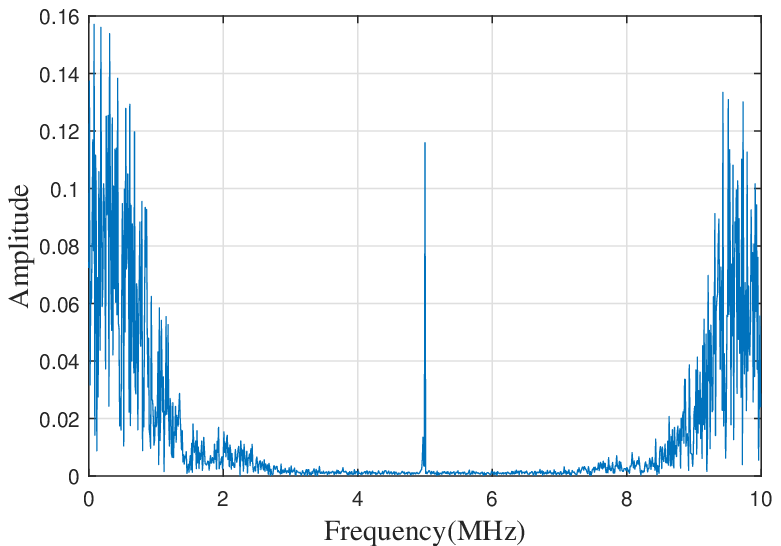}
		}
		\caption{Zigbee signal frequency domain graphs. (a) Indoor-LOS. (b) Indoor-NLOS. (c) Outdoor-LOS. (d) Outdoor-NLOS.}
		\label{sigp52}
	\end{figure*}
	Figure~\ref{sigp51} and \ref{sigp52} display the time-domain and frequency-domain images of a transmitter under four different scenarios. There is a noticeable peak near half of the sampling rate in NLOS conditions, which is typically ignored by CVMD methods. However, the apparent distinctive features can serve as essential bases for identification. Thus, they are input as special features derived from ICVMD into the model, which fully demonstrates the flexibility of the method in its application. Table~\ref{tabzig} lists the number of signal frames and estimated SNR ranges in the four scenarios, where '-' indicates that the signals are transmitted directly, and '+' means the amplifier is added. Each signal frame contains around 250 symbols, and we sliced them into segments with a length of 1000. The training, validation, and test sets were split from different frames in a 4:1:1 ratio to avoid aliasing.
	\begin{table}
		\centering
		\caption{Zigbee Signal Frame Acquisition Distribution \& Estimated SNR Ranges}
		\label{tabzig}
		\begin{tabular}{@{}cccc@{}}
			\toprule
			Scenario &Devices Number& \begin{tabular}[c]{@{}c@{}}Estimated SNR\\Range (dB) \end{tabular} &Frames Number\\ \midrule
			\multirow{2}{*}{Indoor-LOS} & 36+& (34,40)& \multirow{2}{*}{231} \\
			& 24-& (33,38)&\\ \cmidrule(lr){2-3}
			\multirow{2}{*}{Indoor-NLOS}& 36+& (21,31)& \multirow{2}{*}{231} \\
			& 24-& (10,16)&\\ \cmidrule(lr){2-3}
			\multirow{2}{*}{Outdoor-LOS}& 36+& (22,29)& \multirow{2}{*}{114} \\
			& 24-& (13,20)&\\ \cmidrule(lr){2-3}
			\multirow{2}{*}{Outdoor-NLOS} & 36+& (24,30)& \multirow{2}{*}{231} \\
			& 24-& (5,17) &\\ 		
			\bottomrule
		\end{tabular}
	\end{table}
	
	The signal knowledge from the dataset description is helpful in signal processing, which significantly improves the identification. However, to evaluate the framework under real-world non-cooperative conditions, we conducted a blind-test methodology. In this setup, the implementers had no access to any prior information regarding the technical specifications or precise hardware parameters in the dataset, ensuring that the feature extraction and classification were driven solely by the data themselves. Our framework treats the received complex-valued sequences as raw input, employing ICVMD to adaptively estimate signal components under uncertainty. In the experiment, all devices equipped with additional PAs were designated as the identification target. The remaining devices, which share the same signal protocol but possessing distinct hardware characteristics, were utilized as supplements. We implemented rigorous frame-level isolation to eliminate the risk of data leakage: the frames used for training, validation, and testing are entirely disjoint.
	
	We first established a performance baseline using direct classification. The accuracies are presented in Table~\ref{tabfl1}, where columns denote the scenarios and emitters number, and rows are the proportions of training samples. We selected emitters randomly many times and calculated their average accuracy. The proposed method outperforms the conventional one, with apparent differences when using only a few samples. Typically, classification performance decreases as the number of categories increases. However, the results improve when there are very few training samples, because it is equivalent to using more training samples. The experiments also revealed that with sufficient samples, increasing the classifier parameters improved performance on the multi-class classification task, but it is difficult to converge under small samples. To maintain stable results, the model only adjusts the output layer size to match the number of categories, while keeping all other parameters consistent during training.
	\begin{table*}
		\caption{Recognition accuracy of direct classification}
		\label{tabfl1}
		\centering
		\begin{tabular}{@{}cccccccccccccc@{}}
			\toprule
			& &
			\multicolumn{3}{c}{Indoor-LOS} &
			\multicolumn{3}{c}{Indoor-NLOS} &
			\multicolumn{3}{c}{Outdoor-LOS} &
			\multicolumn{3}{c}{Outdoor-NLOS} \\ \midrule
			Accuracy (\%) &
			\begin{tabular}[c]{@{}c@{}}Sample\\ proportion\end{tabular} &
			10 & 20 & 30 & 10 & 20 & 30 & 10 & 20 & 30 & 10 & 20 & 30 \\ \midrule
			\multirow{5}{*}{CNN} & 30\% & 95.7 & 73.2 & 72.3 & 96.4 & 81.1 & 73.6 & 94.9 & 91.6 & 88.2 & 98.0 & 90.9 & 85.8 \\
			& 10\% & 93.7 & 70.1 & 69.8 & 91.9 & 79.2 & 66.3 & 92.9 & 87.6 & 82.9 & 95.4 & 84.6 & 76.4 \\
			& 3\% & 68.5 & 55.9 & 50.1 & 86.2 & 75.8 & 54.4 & 37.6 & 69.9 & 64.5 & 88.4 & 72.3 & 68.4 \\
			& 1\% & 24.8 & 27.6 & 20.4 & 67.4 & 61.2 & 42.1 & 17.8 & 42.9 & 30.8 & 83.0 & 65.9 & 44.3 \\
			 \midrule
			\multirow{5}{*}{\begin{tabular}[c]{@{}c@{}}ICVMD\\ +CNN\end{tabular}} & 30\% & 100 & 93.9 & 90.0 & 98.4 & 89.4 & 82.5 & 98.8 & 94.3 & 89.7 & 99.9 & 91.6 & 86.9 \\
			& 10\% & 100 & 87.0 & 81.7 & 94.6 & 85.0 & 74.0 & 96.1 & 93.3 & 84.6 & 96.2 & 88.4 & 79.3 \\
			& 3\% & 92.8 & 77.1 & 76.1 & 90.2 & 79.5 & 67.6 & 93.2 & 86.2 & 68.9 & 90.9 & 76.5 & 72.3 \\
			& 1\% & 31.3 & 38.0 & 28.0 & 83.4 & 68.8 & 54.9 & 23.5 & 46.8 & 39.6 & 87.7 & 70.3 & 68.1 \\
		 \bottomrule
		\end{tabular}
	\end{table*}
	\subsection{Few-Shot Experiments Analysis}
	\begin{table*}
		\caption{Comparison of few-shot identification methods}
		\label{tabfl2}
		\centering 
		\begin{tabular}{@{}cccccccccccccc@{}}
			\toprule
			& &\multicolumn{3}{c}{Indoor-LOS} & \multicolumn{3}{c}{Indoor-NLOS} & \multicolumn{3}{c}{Outdoor-LOS} & \multicolumn{3}{c}{Outdoor-NLOS} \\ \midrule
			Accuracy (\%)&\begin{tabular}[c]{@{}c@{}}Sample\\proportion\end{tabular}& 10 & 20 & 30& 10 & 20 & 30 & 10 & 20 & 30 & 10& 20 & 30 \\ \midrule
			\multirow{4}{*}{MAML \cite{MAML}} &10\%& 95.6 & 76.5 & 73.7& 93.2 & 85.0 & 79.4 & 93.1 & 88.8 & 85.1 & 96.8& 88.3 & 83.6 \\
			&3\%& 86.0 & 67.5 & 61.5& 88.2 & 81.2 & 73.5 & 88.5 & 81.3 & 74.0 & 91.1& 80.3 & 79.1 \\
			&1\%& 82.2 & 62.1 & 54.6& 82.2 & 76.4 &63.1 & 81.9 & 80.7 & 68.2 & 87.7& 75.0 & 67.5 \\
			 \midrule
			\multirow{4}{*}{\begin{tabular}[c]{@{}c@{}}Proposed\\ (Basic)\end{tabular}} &10\% & 100& 96.0 & 92.3& 96.2 & 92.5 & 84.2& 97.2 & 94.2 & 87.2 & 97.9& 93.6 & 90.3 \\
			&3\%& 93.4 & 84.7 & 83.4& 91.6 & 85.3 & 76.3 & 94.5 & 88.7 & 77.1 & 94.4& 86.5 & 86.7 \\
			&1\%& 91.9 & 80.8 & 75.3& 86.3 & 76.9 & 66.8 & 87.2 & 84.0 & 74.4 & 88.1& 78.2 & 77.0 \\
			 \midrule
			\multirow{4}{*}{InterML \cite{IntML}}  & 10\% & 99.9 & 99.3 & 89.1 & 95.7 & 92.7 & 86.4 & 97.4 & 94.4 & 92.8 & 96.0 & 94.9 & 92.7 \\
			& 3\% & 95.4 & 91.2 & 82.0 & 89.3 & 86.0 & 79.3 & 92.1 & 85.9 & 81.9 & 93.9 & 90.6 & 87.9 \\
			& 1\% & 86.8 & 73.2 & 63.6 & 81.5 & 78.4 & 67.7 & 85.3 & 73.5 & 70.8 & 89.0 & 81.5 & 75.7 \\
			\midrule
			\multirow{4}{*}{\begin{tabular}[c]{@{}c@{}}Proposed\\(Diverse)\end{tabular} }&10\%& 99.9 & 96.7 & 94.6& 99.7 &93.4 & 88.5 & 97.8 & 95.7 & 92.4 & 100 & 96.0 & 94.5 \\
			&3\%& 99.7 & 96.2 & 87.4& 96.3 & 87.2 & 81.6 & 95.9 & 94.4 & 89.6 & 99.7& 95.4 & 93.7 \\
			&1\%& 97.5 & 93.0 & 83.9& 87.4 & 78.5 & 70.3 & 89.3 & 88.6 & 86.5 & 97.7& 91.4 & 91.1 \\
		\bottomrule
		\end{tabular}
	\end{table*}
	The proposed method was evaluated against established FS approaches, including MAML \cite{MAML} and InterML \cite{IntML,BIM}, leveraging the auxiliary data. As shown in Table~\ref{tabfl2}, the performance is substantially improved across all scenarios compared to the direct classification. When the training proportion is only 1\%, our method achieves at least 66.8\% (basic) and 70.3\% (diverse) accuracy for 30 categories, and single-run minima are close to or exceed the previous best results. Additionally, standard deviations for the key 1\% training proportion entries are reported as an additional row in the table. Our method achieves the lowest standard deviation, and this superior stability stems from the transfer learning mechanism. Conversely, InterML method exhibits a wider variance owing to uncontrolled synthetic sample quality.
	
	We observed that signal variations are more pronounced between different frames than within a single frame. In previous experiments (basic), when reducing the training samples, we prioritized frame reduction and then decreased the symbols per frame for extremely low sample ratios to fit reality. This ensures the training set encompasses sufficient inter-frame diversity, preventing the model from overfitting to specific captures and providing a reliable assessment of the FS capability. We explored different FS setting strategy to provide a comprehensive evaluation and strictly controlled the total volume of training data. We shifted from using a few full-length frames to a synchronized reduction strategy, where both frame number and the symbols per frame are adjusted. This approach  allows collecting shorter segments from more independent frames, ensuring that the training set is sourced from a broader range of independent captures. The updated strategy (diverse) yielded the performance improvements shown in Table~\ref{tabfl2}, confirming that inter-frame signal variations are greater than those between segments within the same frame. Training on more independent frames yields more discriminative emitter features.
	
	\begin{figure*}
		\centering
		\subfloat[]{		
			\includegraphics[width=0.245\linewidth]{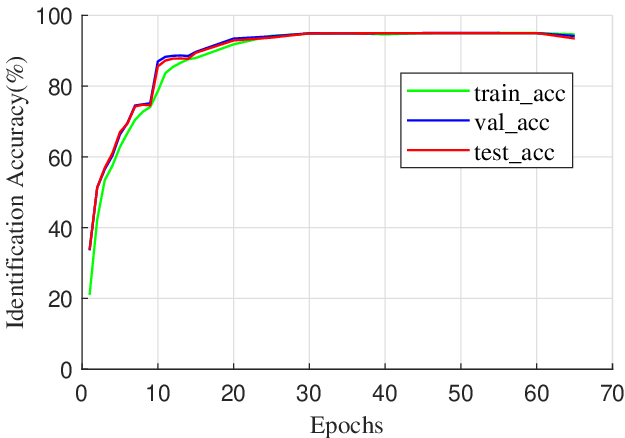}
		}
		\subfloat[]{		
			\includegraphics[width=0.245\linewidth]{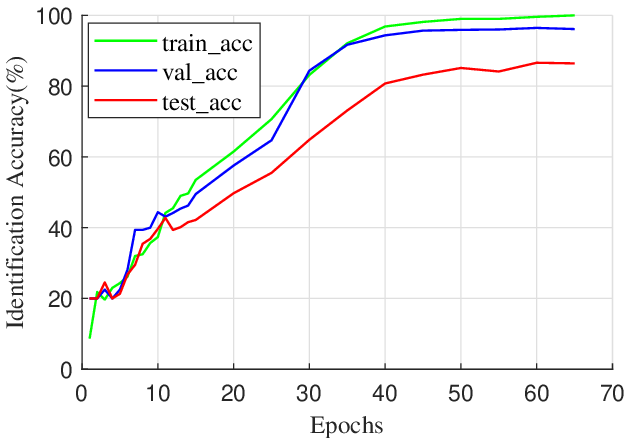}
		}
		\subfloat[]{	
			\includegraphics[width=0.245\linewidth]{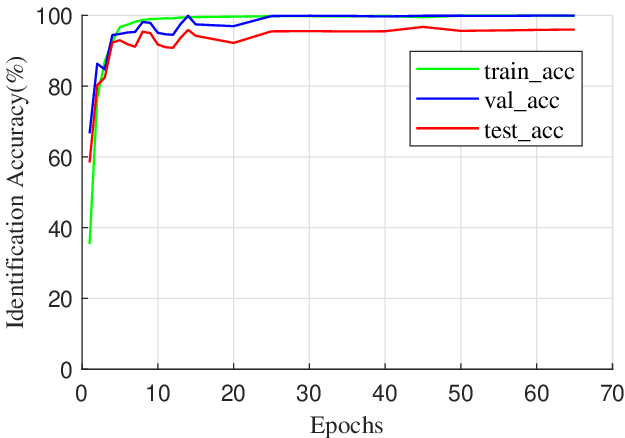}
		}
		\subfloat[]{
			
			\includegraphics[width=0.245\linewidth]{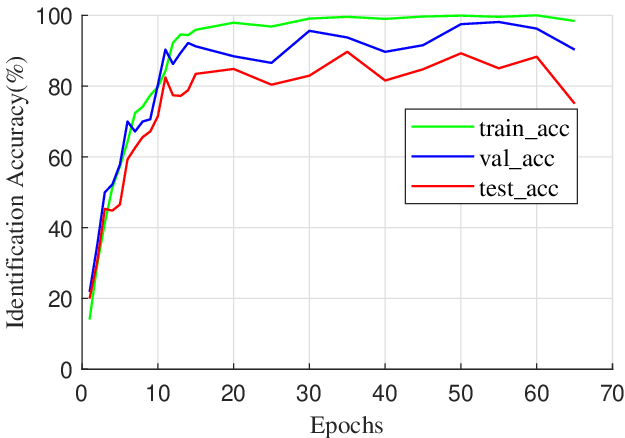}
		}
		\caption{Accuracy trend figures during recognition in four scenarios. (a) Indoor-LOS. (b) Indoor-NLOS. (c) Outdoor-LOS. (d) Outdoor-NLOS.}
		\label{sigp53}
	\end{figure*}
	To further investigate model behavior, we monitored the training process of the ICVMD+CNN model and plotted the recognition curves across four scenarios. Fig.~\ref{sigp53} reveals that validation and test accuracies are essentially consistent in the indoor-LOS scenario, while discrepancies with training accuracy persist. In other scenarios, the three curves are distinctly separated. Especially in the indoor-NLOS scenario, the training accuracy converges above 99\% while the validation accuracy is around 96\%, but the test accuracy never reaches 90\%. This clearly indicates overfitting caused by these significant inter-frame signal variations.
	
	\begin{figure*}
		\centering
		\subfloat[]{		
			\includegraphics[width=0.245\linewidth]{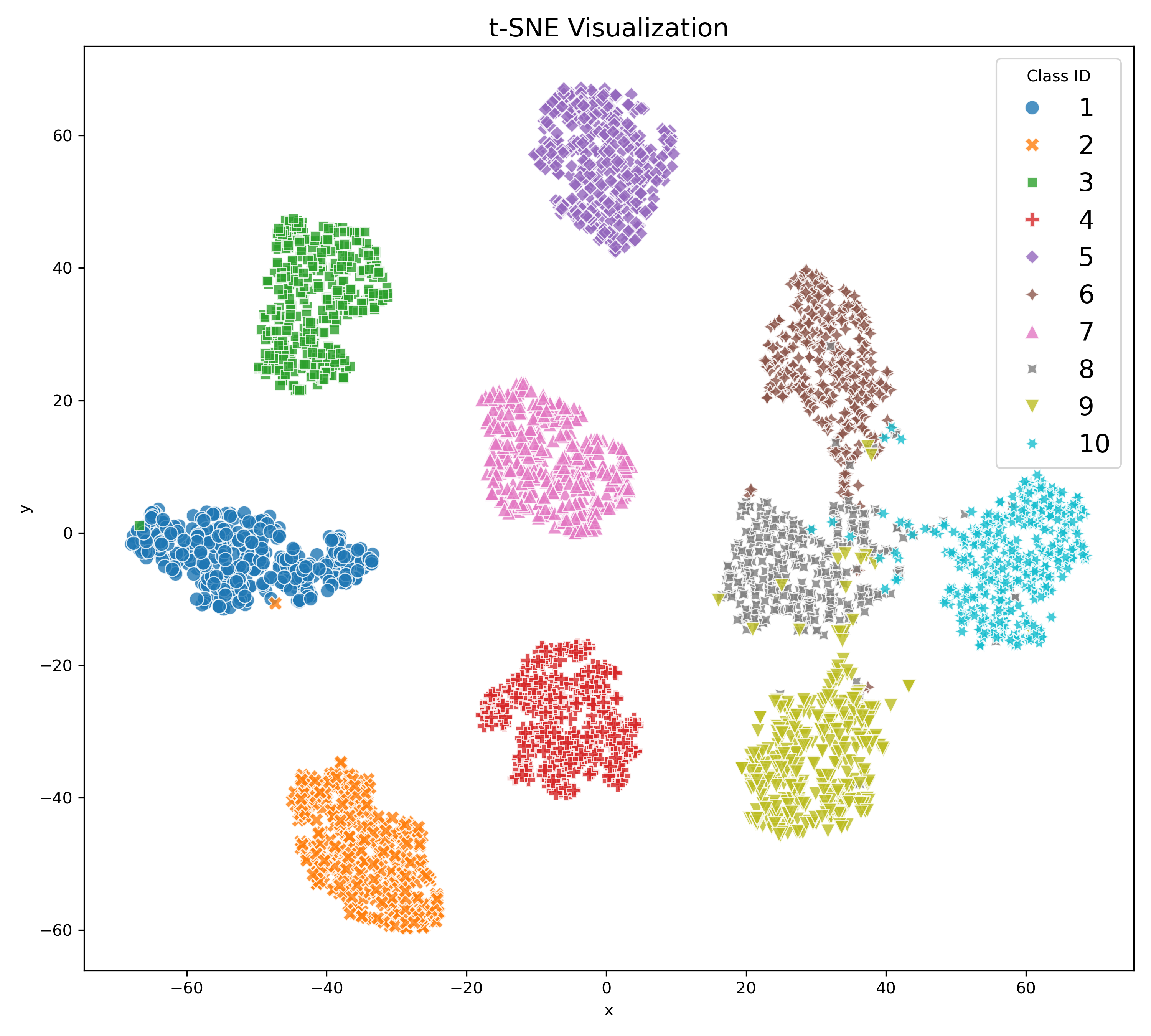}
		}
		\subfloat[]{		
			\includegraphics[width=0.245\linewidth]{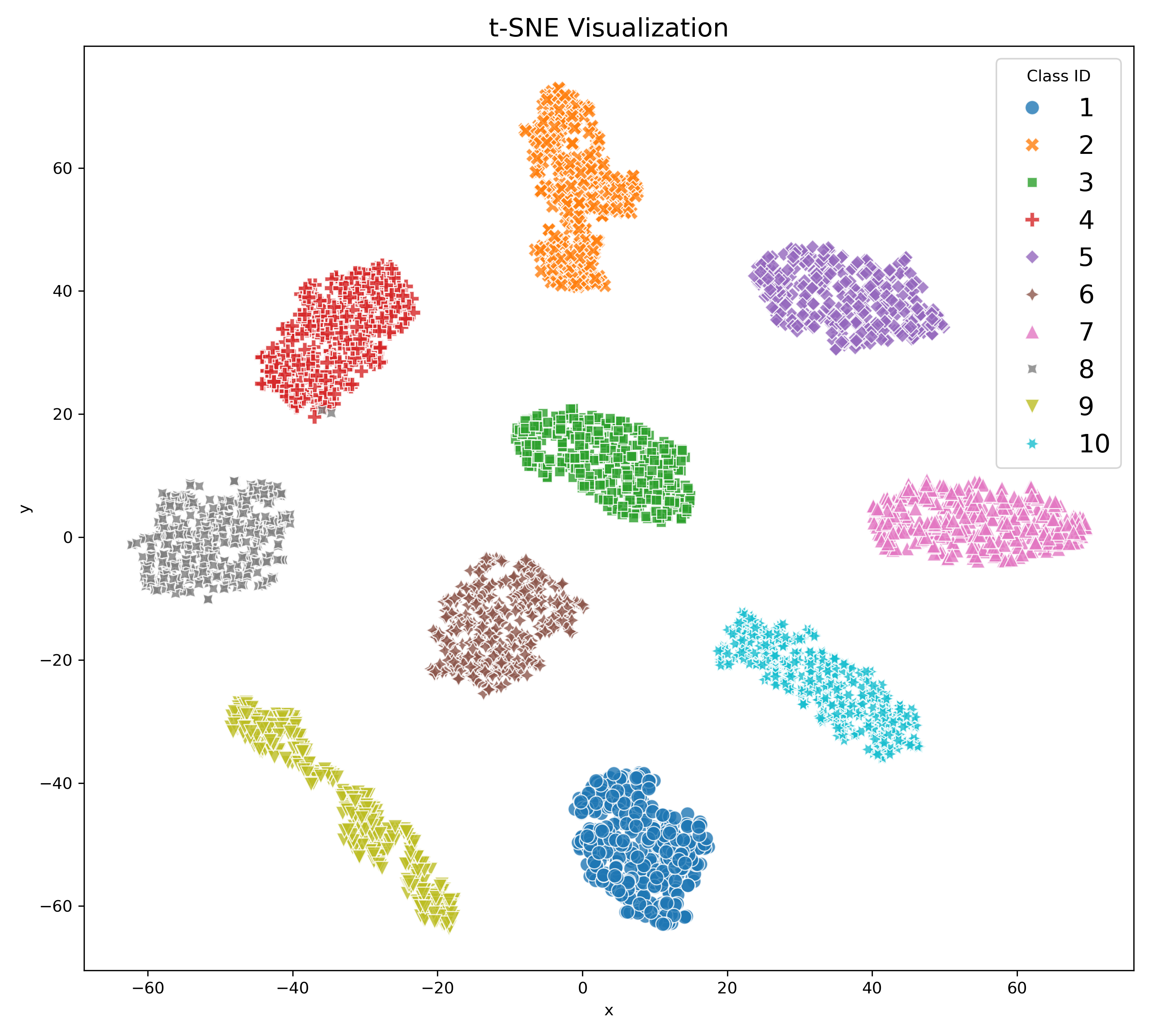}
		}
		\subfloat[]{	
			\includegraphics[width=0.245\linewidth]{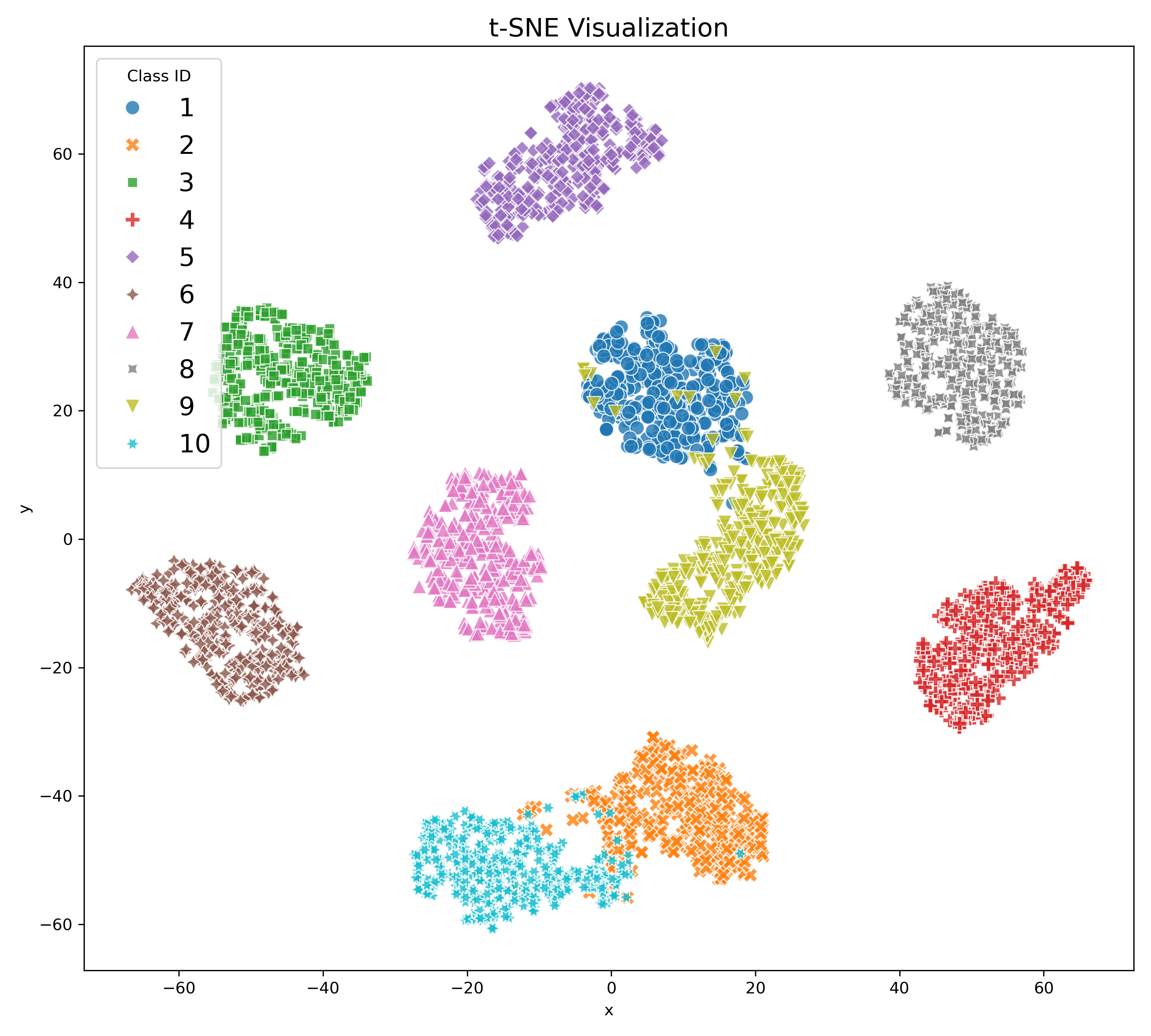}
		}
		\subfloat[]{
			\includegraphics[width=0.245\linewidth]{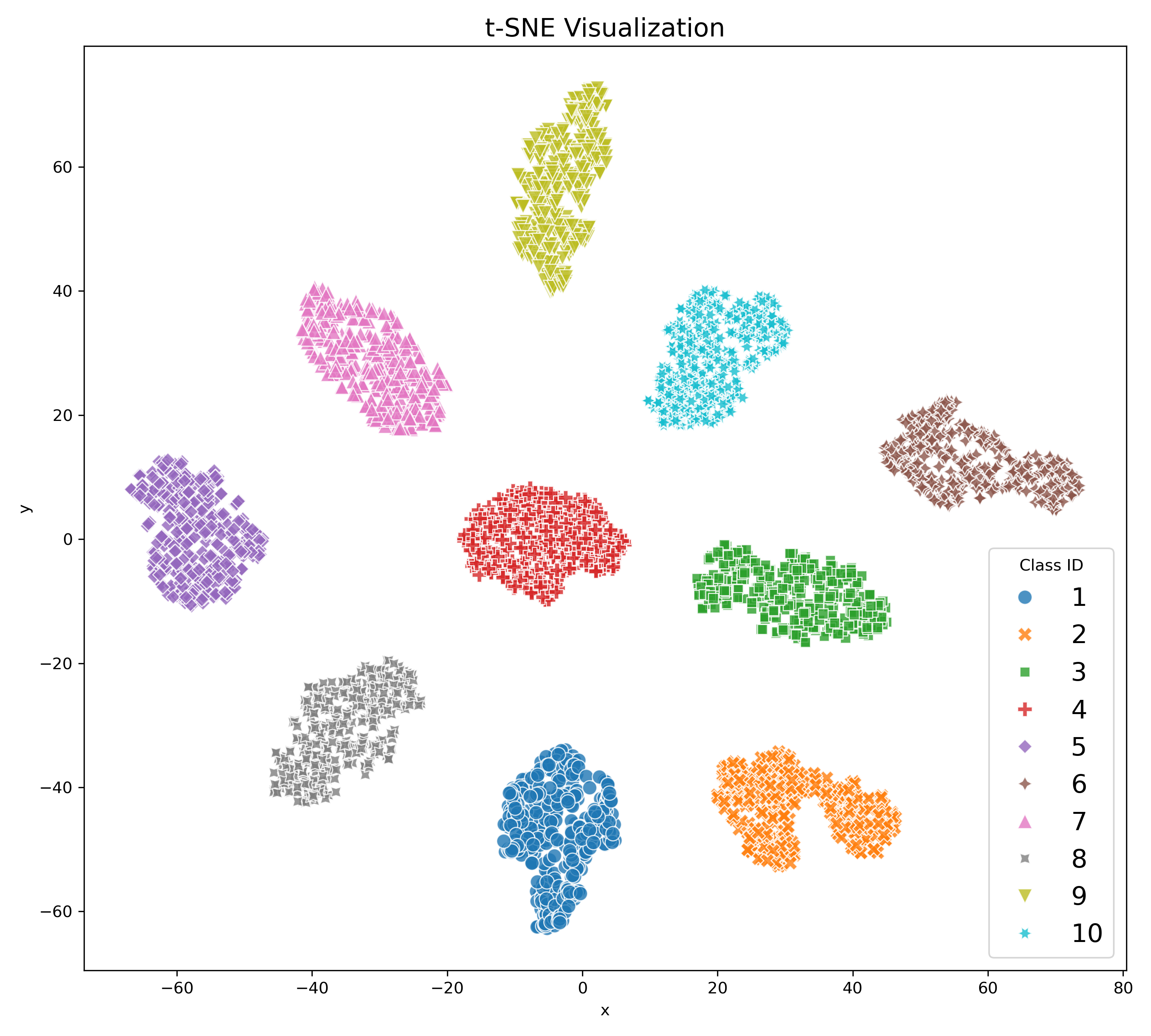}
		}
		\caption{t-SNE visualization of the features in four scenarios (10 classes). (a) Indoor-LOS. (b) Indoor-NLOS. (c) Outdoor-LOS. (d) Outdoor-NLOS.}
		\label{sigp55}
	\end{figure*}
	To evaluate the model's feature extraction capability, we utilized t-SNE to visualize the high-dimensional feature distributions. As illustrated in Fig.~\ref{sigp55}, the features for 10 emitters are mapped onto a 2D space across four scenarios. The plots reveal that the proposed method produces highly concentrated intra-class clusters and maintains inter-class distances. These visualizations support the quantitative results in Table~\ref{tabfl2}, demonstrating that the architecture can extract stable and discriminative hardware characteristics.
	
	\begin{figure*}
		\centering
		\subfloat[]{		
			\includegraphics[width=0.45\linewidth]{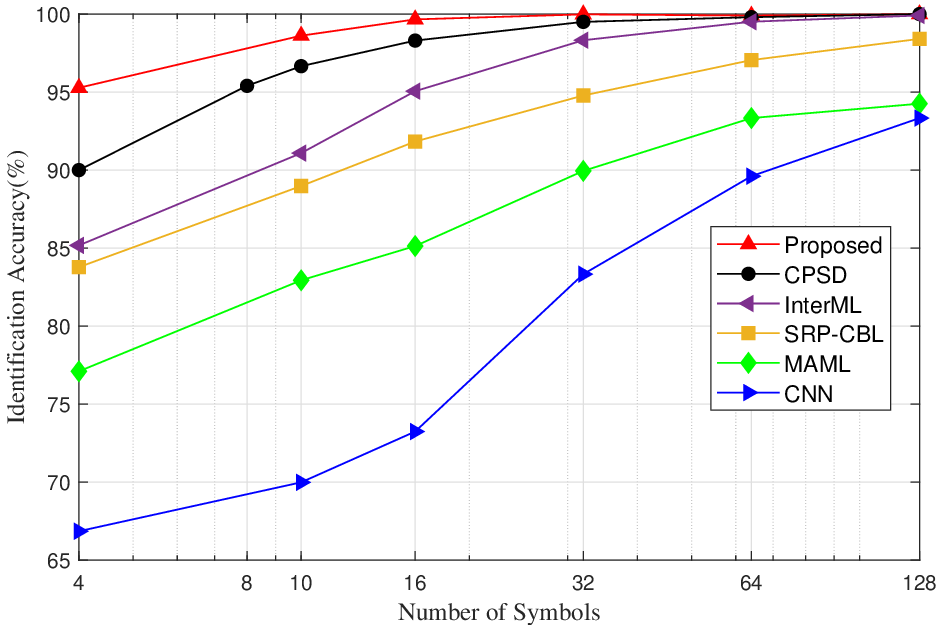}
		}\hfil
		\subfloat[]{		
			\includegraphics[width=0.45\linewidth]{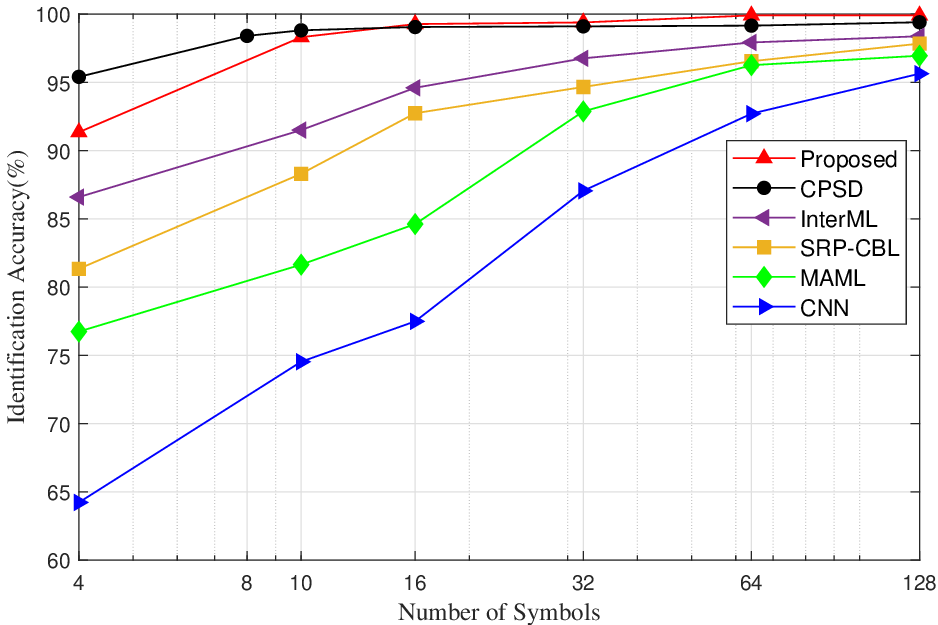}
		}
		\\
		\subfloat[]{	
			\includegraphics[width=0.45\linewidth]{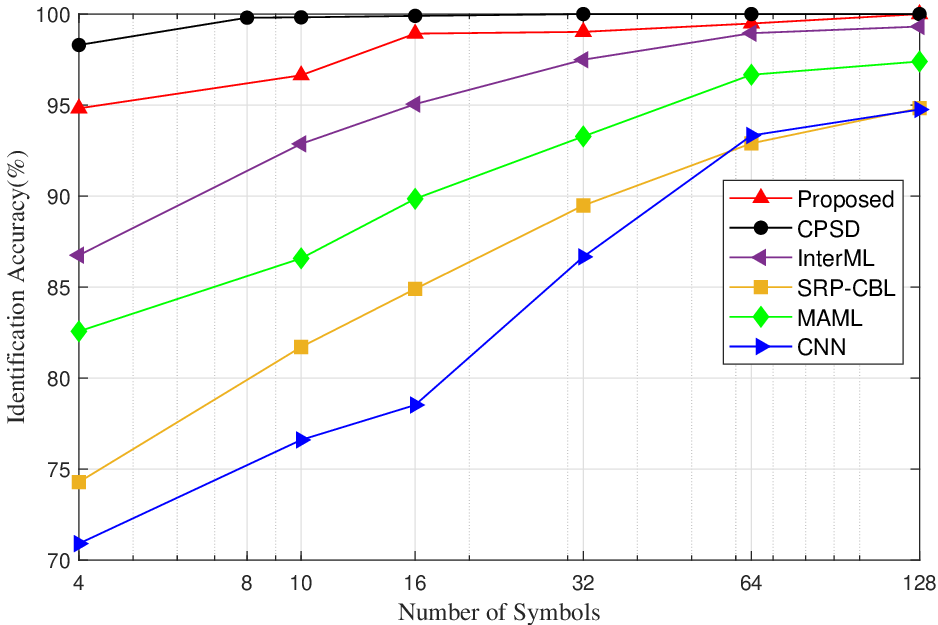}
		}\hfil
		\subfloat[]{
			
			\includegraphics[width=0.45\linewidth]{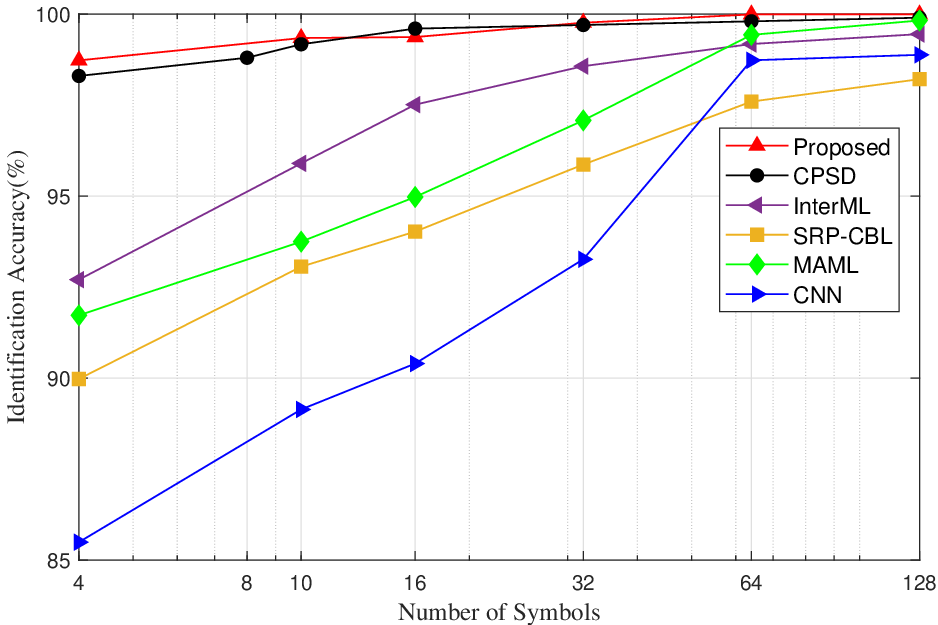}
		}
		\caption{Comparison of recognition accuracy in four scenarios. (a) Indoor-LOS. (b) Indoor-NLOS. (c) Outdoor-LOS. (d) Outdoor-NLOS.}
		\label{sigp54}
	\end{figure*}
	Finally, we benchmark against the SRP-CBL \cite{SRP} and cross power spectral density (CPSD) algorithm \cite{cycle}, the latter of which is the current state-of-the-art (SOTA) method specifically designed for this dataset. There is a difference in experimental setups, so we aligned our settings to maintain close conditions and compare directly with the reported reliable results. Since both training and validation sets are available, there is no need to enforce a strict frame-level isolation between them, while the test frames are strictly disjoint from both training and validation frames throughout. Within the training split, segments are randomly shuffled, and a portion is held out as the validation set, which is used exclusively for early stopping and model selection. No segments overlap between the training, validation, and test subsets. As illustrated in Fig.~\ref{sigp54}, our proposed method and CPSD form the top-performing tier, distinctly outperforming other approaches. Without using prior signal information (which CPSD mainly relies on for signal processing), our method closely matches its performance and even surpasses it in the indoor-LOS scenario. The accuracy of SRP-CBL drops in outdoor scenarios, sometimes falling below baseline. However, even with this performance fluctuation, SRP-CBL and all other methods consistently outperform the traditional CNN under FS conditions. This gap widens as the number of symbols decreases, demonstrating the effectiveness and necessity of methods designed explicitly for FS scenarios.
	
	Despite the superior performance in FS regimes, the proposed framework has certain limitations. First, while ICVMD offers excellent adaptability in non-cooperative scenarios where signal parameters are unknown, its performance upper bound may be slightly lower than that of specialized signal processing methods when precise prior information is available. Second, as an adaptive decomposition approach, the stability of ICVMD can be affected by extreme signal variations, requiring more robust parameter initialization. Finally, the integration of TCN and SAM modules increases the architectural coordination required during training, though the additional inference overhead is acceptable. Future work will focus on optimizing collaboration and improving the stability of adaptive decomposition under dynamic interference.
	\section{Conclusion}\label{sec5}
	In this paper, we have proposed a novel FS-SEI method with signal decomposition and SAT to overcome target data limitations. We extended the VMD to the complex domain and recovered the original signals through recombining IMFs, thereby constructing connections between signals as the source of RFFs. The RFF extraction and classification were performed in an end-to-end manner, and we employed a multi-scale TCN for effective sequential feature processing. Given the small sample size, the FCN classifier helps prevent overfitting. We utilized the signal-guided spatial attention network trained on the auxiliary dataset to achieve accurate attention determination. Ablation experiments on simulated data demonstrated the positive contribution of each component to overall recognition performance. Through experiments conducted with public real-world data, our approach achieves comparable performance despite lacking critical signal information compared to existing methods. In summary, the ICVMD-SAT method effectively extracts RFFs and demonstrates strong identification performance across both simulated and four real-world indoor/outdoor conditions. In the future, we will consider scenarios with greater domain differences, like variable modulation identification, to further enhance the transfer ability when the source and target domains differ.
	
	\bibliographystyle{IEEEtran}
	\bibliography{cs123}
	\end{document}